\def\etal{{\it et al.}\thinspace}
\def\ie{{\it i.e.,}\thinspace}
\def\eg{{\it e.g.,}\thinspace}
\documentclass[useAMS]{mnras}
\usepackage{graphicx}
\usepackage{rotating}
\usepackage{subfigure}
\usepackage{multirow}
\usepackage{verbatim}
\usepackage{amssymb}
\def\eq{\begin{equation}}
\def\en{\end{equation}}

\def\deg{\hbox{$^{\circ}$}}

\def\aj{{\it A.J.}\thinspace}
\def\apj{{\it Ap.J.}\thinspace}
\def\apjl{{\it Ap.J. Lett.}\thinspace}
\def\apjs{{\it Ap.J. Suppl.}\thinspace}
\def\aap{{\it A\&A}\thinspace}

\def\mnras{{\it MNRAS}\thinspace}
\def\P3hat{{\mathaccent 94 P}_3}

\title[Core and Conal Component Analysis of Pulsar B1933+16]{Core and Conal Component Analysis of Pulsar B1933+16 --- Investigation of the Segregated Modes}
\author[Dipanjan Mitra, Joanna Rankin \& Mihir Arjunwadkar]{Dipanjan Mitra$^{1,2,3}$Joanna Rankin$^{1,4}$ \& Mihir Arjunwadkar$^{5}$ \\
$^1$Physics Department, University of Vermont, Burlington, VT 05405\thanks{dmitra@uvm.edu; Joanna.Rankin@uvm.edu} \\
$^2$National Centre for Radio Astrophysics, Ganeshkhind, Pune 411 007 India\thanks{dmitra@ncra.tifr.res.in} \\
$^3$Janusz Gil Institute of Astronomy, University of Zielona G\'ora, Lubuska 2, 65-265 Zielona G\'ora, Poland\\
$^4$Sterrenkundig Instituut `Anton Pannekoek', University of Amsterdam, NL-1090 GE\\ 
$^5$Centre for Modeling and Simulation, Savitribai Phule Pune University, Ganeshkhind, Pune 411 007 India\thanks{mihir@cms.unipune.ac.in}}

\date{In original form year month day}
\pubyear{year}
\begin{document}
\label{firstpage}
\pagerange{\pageref{firstpage}--\pageref{lastpage}}
\maketitle

\begin{abstract}
Radio pulsar B1933+16 is brightest core-radiation dominated pulsar in
the Arecibo sky, and here we carry out a comprehensive high resolution
polarimetric study of its radiation at both 1.5 and 4.6 GHz.  At 1.5
GHz, the polarization is largely compatible with a rotating-vector
model with $\alpha$ and $\beta$ values of 125 and --1.2\degr, such
that the core and conal regions can be identified with the primary and
secondary polarization modes and plausibly with the extraordinary and
ordinary propagation modes.  Polarization modal segregation of
profiles shows that the core is comprised of two parts which we
associate with later X-mode and earlier O-mode emission.  Analysis of
the broad microstructures under the core shows that they have similar
timescales to those of the largely conal radiation of other pulsars
studied earlier.  Aberration/retardation analysis was here possible
for both the conal and core radiation and showed average physical
emission heights of about 200 km for each.  Comparison with other
core-cone pulsars suggests that the core and conal emission arises
from similar heights.  Assuming the inner vacuum gap model, we note
that at these emission altitudes the frequency of the observed
radiation $\nu_{obs}$ is less than the plasma frequency $\nu_p$.  We
then conclude that the radio emission properties are consistent with
the theory of coherent curvature radiation by charged solitons where
the condition $\nu_{obs} < \nu_{p}$ is satisfied.  However, the
differences that exist between core and conal emission in their
geometric locations within a pulse, polarization and modulation
properties are yet to be understood.

\end{abstract}


\maketitle

\section{Introduction}
\label{sec1}
Radio pulsar B1933+16 is the brightest core-emission dominated pulsar
in the Arecibo sky, as the majority of such pulsars fall far in the
southern sky around the Galactic center.  Despite its prominence, it
has received very little study and analysis.  Sieber \etal\ (1975)
early showed its profile evolution from a single form at meter
wavelengths to a tripartite one above 1 GHz, and this pattern of
evolution later became a defining characteristic of the core-single
profile class in the {\it Empirical Theory of Pulsar Emission} series
(Rankin 1983, 1993a; hereafter ET I, VIa).  Other work (Gould \& Lyne
1998; Rankin, Stinebring \& Weisberg 1989; Weisberg \etal 1999;
Hankins \& Rankin 2010) confirmed this profile evolution with even
better quality observations.

In the parlance of the ET papers, B1933+16 has been regarded as having
a core-single ({\bf S}$_t$) profile, meaning that it ``grows'' conal
``outriders'' only at high frequency.  The star's core feature does
exhibit the antisymmetric circular polarization often seen under core
features, as we see in Fig.~\ref{fig1} below, and the angular
dimensions of the core and conal components seem to reflect the
angular size of the pulsar's polar cap emission region as expected in
ET VI.  However, B1933+16's core ``component'' does exhibit a very
clear non-Gaussian structure---as do some other core features
(sometimes earlier referred to as ``notches'')---and its origin is one
of the questions before us below.  At lower frequencies its core
component gradually broadens (ET II), in part because of scattering,
and by 100 MHz, the scattering tail occupies the entire rotation cycle
(Izvekova \etal\ 1989).

Few pulse-sequence analyses had been carried out on pulsars with
dominant core emission.  Exceptions are B0329+54 and the Vela pulsar
B0833--45, the study of this latter pulsar by Krishmohan \& Downs
(1983) being an early classic which has received far too little
followup.  Our study of B0329+54's normal mode (Mitra, Rankin \& Gupta
2007) revealed a clear signature of intensity-dependent
aberration/retardation (hereafter A/R) in its core emission, which in
turn was interpreted as the presence of a linear amplifier within the
star's polar flux tube.  A further study showing similar core
properties was conducted with pulsar B1237+25 (Smith, Rankin \& Mitra
2012).

Johnston \etal\ (2005) first demonstrated that the ``fiducial''
polarization orientation of pulsar radio emission---at the central
longitude of the magnetic axis---exhibits preferential parallel or
orthogonal alignments with respect to their proper-motion (hereafter
$PA_v$) directions.  In carrying this work further by considering only
pulsars with core emission, Rankin (2015) showed that the later core
emission shows an accurate orthogonal orientation reflecting the
extraordinary X propagation mode.  The classic Johnston \etal\ study
did include pulsar B1933+16; but, in their analysis its alignment was
indifferent.  We have come to believe, however, that the longitude of
the pulsar's magnetic axis falls later than was then thought, and when
this is taken into account, B1933+16's late core emission seems to be
the X mode and, if so, the two polarization modes can be identified
throughout its profile.  This will be further discussed below.

In this current analysis we have a few straightforward objectives that
follow from our earlier pulse-sequence analyses of core emission
above: a) we want to assess whether the star's core radiation exhibits
intensity-dependent A/R and, if so, interpret its physical
significance; b) we will attempt to understand how the pulsar's
polarization position-angle (hereafter PPA) traverse becomes distorted
from what would be expected from the rotating-vector model (hereafter
RVM; Radhakrishnan \& Cooke 1969); we want to understand the
significance of the pulsar's prominent circular polarization; further,
B1933+16 is one of the few bright pulsars available to us in which
core microstructure can be investigated, so we will attempt an
analysis similar to that of Mitra, Arjunwadkar \& Rankin (2015); and
finally we will assess how the pulsar's conal emission is connected to
its prominent core radiation.

\begin{figure}
\begin{center}
\mbox{\includegraphics[height=100mm,width=75mm,angle=-90.]{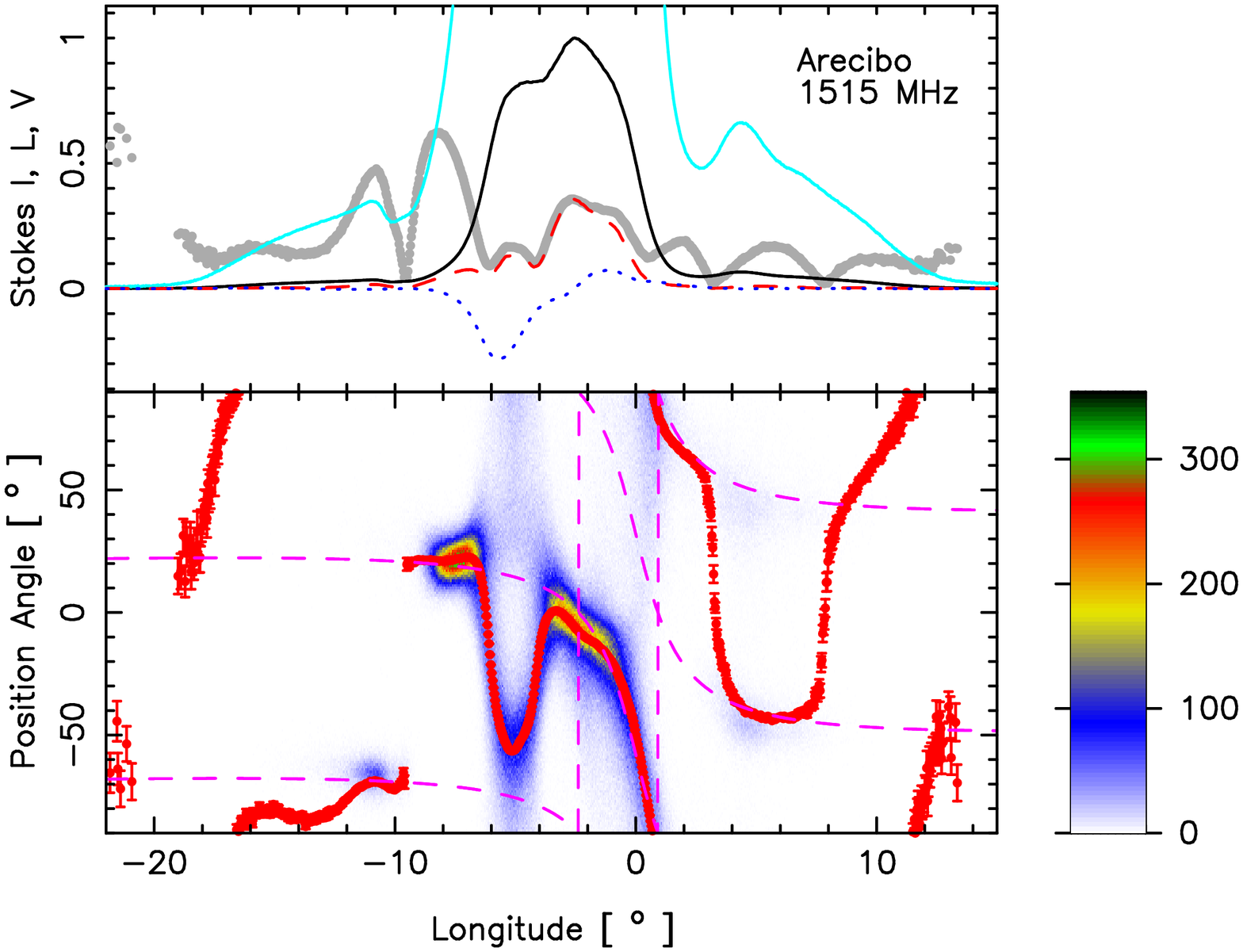}}
\mbox{\includegraphics[height=100mm,width=75mm,angle=-90.]{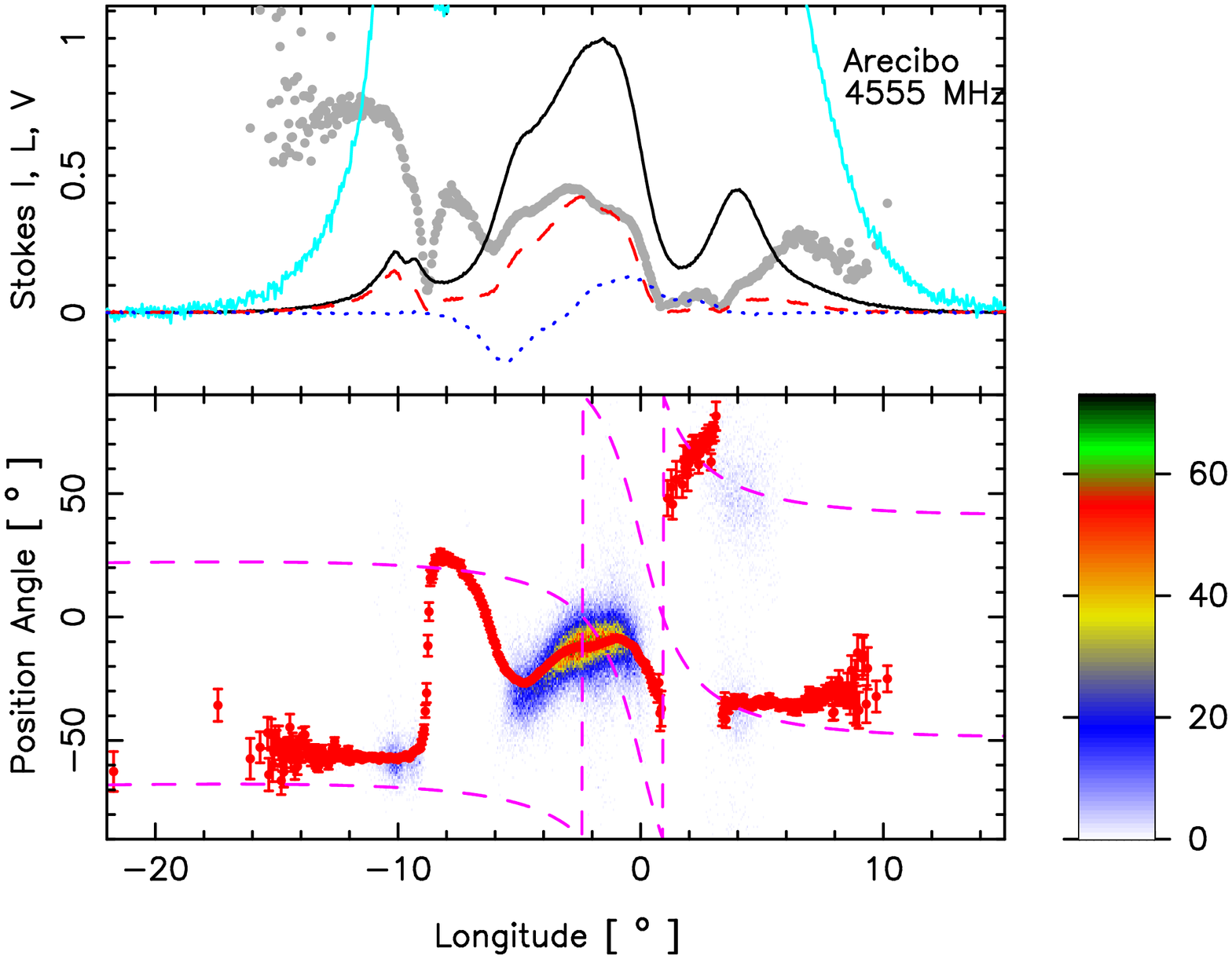}}
\caption{ Total polarization displays at 1515 MHz (top) and 4555 MHz (bottom), 
giving (upper panels) the total intensity (Stokes $I$; solid curve),
the total linear ($L$ (=$\sqrt{Q^2+U^2}$); dashed red), the circular
polarization (Stokes $V$; dotted blue), the fractional linear
polarization ($L/I$; gray points) and a zoomed total intensity
(10$\times I $; cyan curve).  The PPA [=${1\over 2}\tan^{-1} (U/Q)$]
densities (lower panels) within each 1$\degr$x1-sample cell correspond
to samples larger than 2 $\sigma$ in $L$, are plotted according to the
color bars at the lower right, and have been derotated to infinite
frequency.  The average PPA traverses (red) are overplotted, and the
RVM fit to the 1515-MHz PPA traverse is plotted twice for the two
polarization modes (magenta dashed) on each panel.  The origin is
taken at the fitted PPA inflection point.}
\label{fig1}
\end{center}
\end{figure}

In what follows, \S\ref{sec2} describes our observations. \S\ref{sec3}
presents our polarimetry, and \S\ref{sec4} discusses the results of
intensity fractionation.  In \S\ref{sec5} we conduct analyses of the
segregated modes, \S\ref{sec6} interprets the PPA traverse
geometrically, and \S\ref{sec7} presents the microstructure
analysis.  \S\ref{sec8} discusses the core and conal emission,
\S\ref{sec9} provides an aberration/retardation analysis, and \S\ref{sec10} 
considers the frequency independence of the modal polarization.  
\S\ref{sec:summary} then summarizes our results and \S\ref{sec:discussion} 
considers their larger theoretical interpretation and significance.

\section{Observations}
\label{sec2}
The observations used in our analyses were made using the 305-m
Arecibo Telescope in Puerto Rico.  The primary 1515-MHz polarized
pulse sequence (hereafter PS) was acquired with the upgraded
instrument using seven Mock Spectrometers on 2015 April 7 (MJD 57119)
and is comprised of 5017 pulses.  Each spectrometer sampled a 25-MHz
portion of the 275-MHz band centered at 1515 MHz with 36.14 $\mu$s
sampling.  The 4555-MHz observation was again carried out using seven
Mocks on 2015 September 20 (MJD 57284) with a total bandwidth of 7x150
MHz and 51 $\mu$s sampling.  The Stokes parameters have been corrected
for dispersion, interstellar Faraday rotation and various instrumental
polarization effects and are ultimately derotated to infinite
frequency.  Errors in the PPAs were computed relative to the off-pulse
noise phasor---that is, $\sigma_{PPA}
\sim tan^{-1 } (\sigma_{off-pulse}/L)$.

\begin{figure}
\begin{center}
\begin{tabular}{c}
\mbox{\includegraphics[height=78mm,angle=-90.]{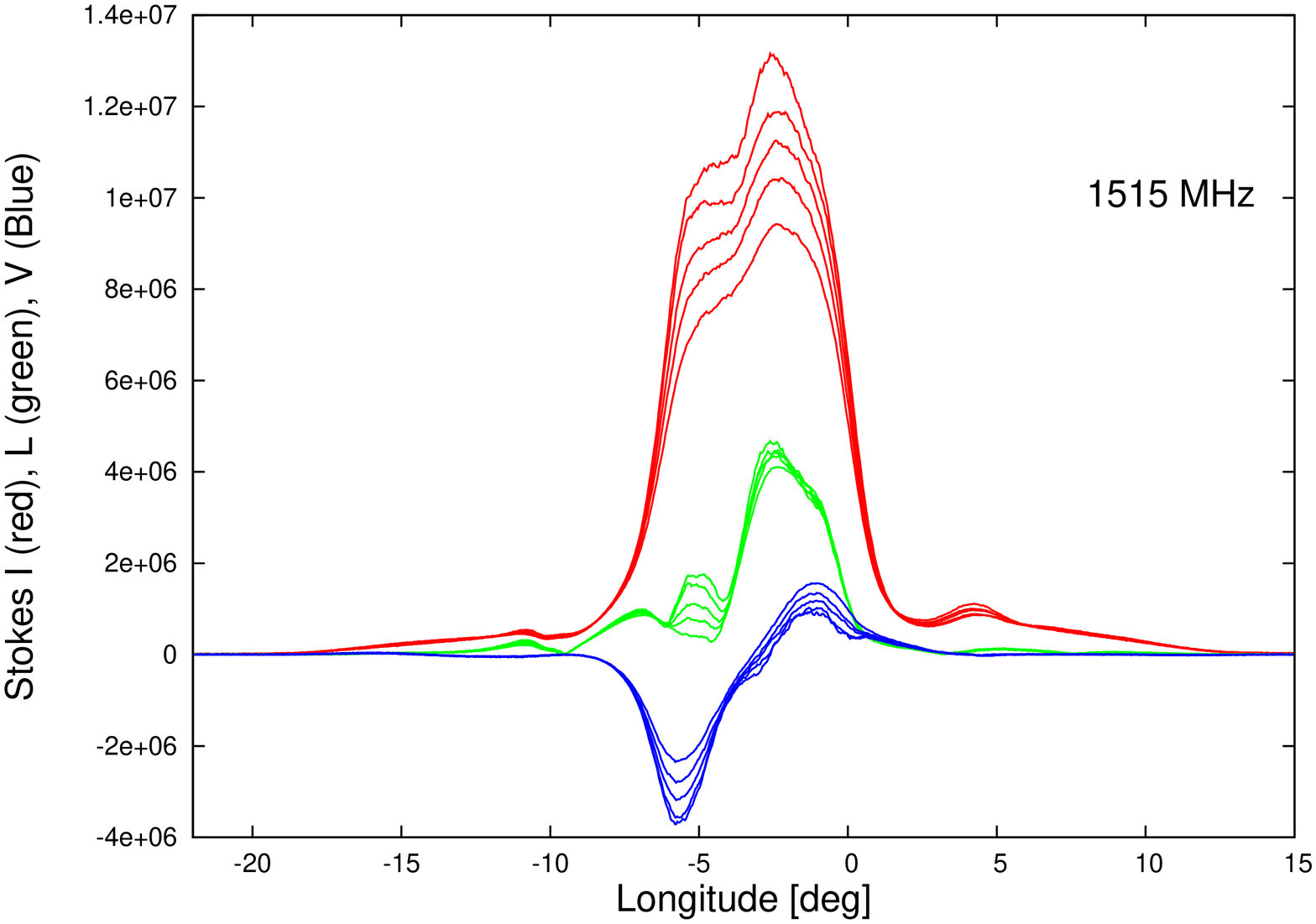}}\\
\mbox{\includegraphics[height=78mm,angle=-90.]{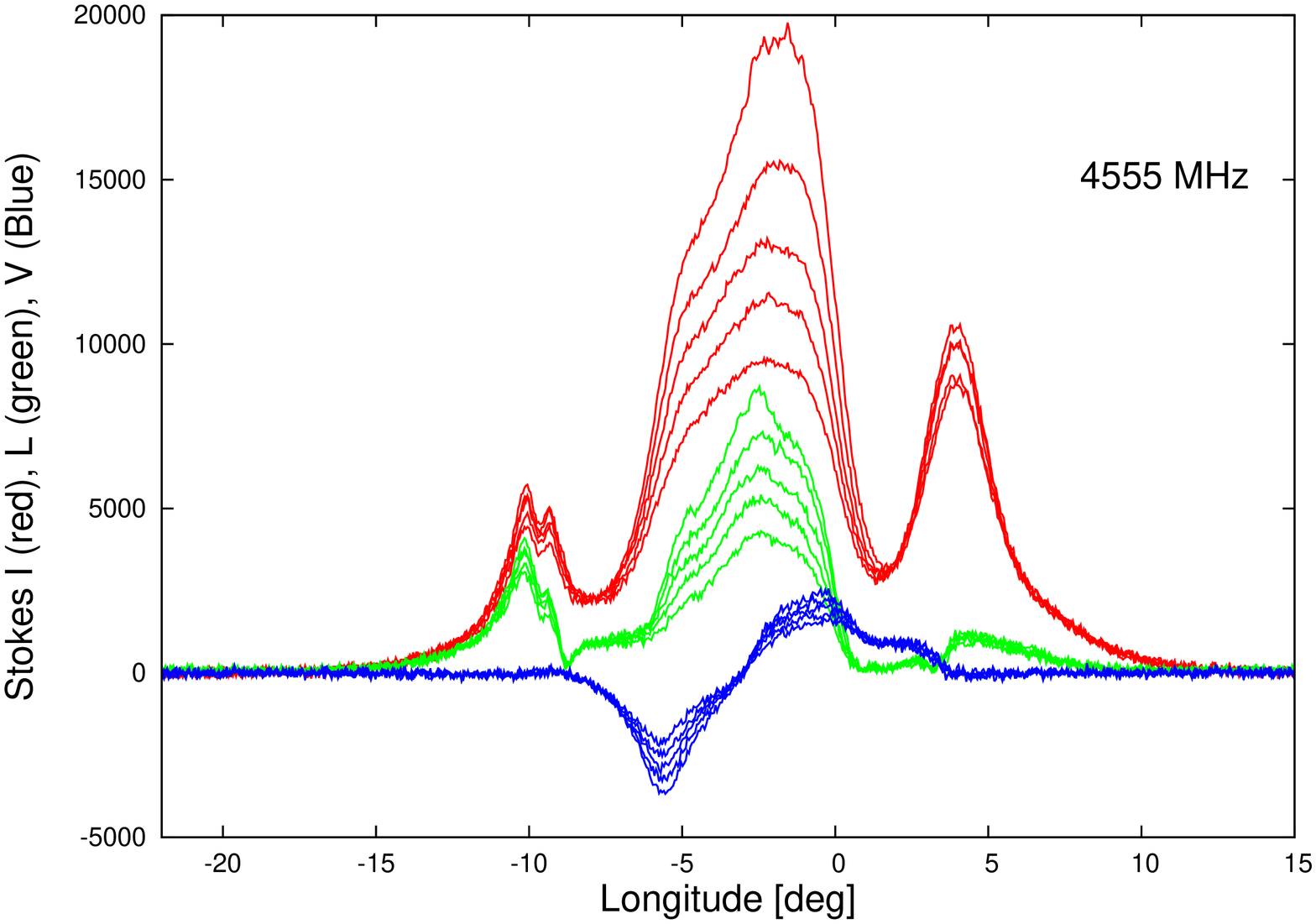}}\\
\end{tabular}
\caption{Average profiles corresponding to five intensity fractions for the 1.5-GHz 
(upper) and 4.6-GHz (lower) observations.  Overall, this analysis
reflects the steadiness of the pulsar's emission.  However, several
aspects are worthy of mention.  Note the manner in which the leading
part of the core becomes progressively weaker relative to the trailing
region with increased intensity.  Also, note that this leading part
becomes more linearly polarized.}
\label{fig2}
\end{center}
\end{figure}

\section{Confronting the Polarimetry}
\label{sec3}
The average 1515- and 4555-MHz polarimetry presented in
Fig.~\ref{fig1} is new and unique in several ways.  First, the
pulsar is strong enough that a large proportion of the respective 36
and 51-$\mu$s samples show well defined polarization---and the linear
polarization angle (hereafter PPA) histograms corresponding to them
are plotted in the lower panels along with the average PPA
traverse. The histograms also have the effect of revealing the
polarized structure of the 1.5-GHz PPA ``notch'' at --5\degr\ much
more clearly, as in earlier profiles it is seen only a ``bump'' in the
average PPA traverse.  The PPAs have been derotated to infinite
frequency, so that it is easy to see that several ``patches'' of PPAs
at the two frequencies correspond to each other.

The puzzling structure of the central core component is clear in the
total power profiles at both frequencies (solid black curves), and
while earlier noted as a ``notch'' possibly due to ``absorption'', we
will explore whether it has a different origin.  The aggregate linear
polarization (dashed red) is only moderate at both frequencies across
the full width of the profile, and we see a consistent signature of
circular polarization (dotted blue).  The conal outriding components
are more prominent at the higher frequency as expected, and PPA
``patches'' corresponding to their orthogonal polarization modes
(hereafter OPMs) fall at similar positions in the respective lower
panels.

Only at the lower frequency is it possible to fit a rotating-vector
model (hereafter RVM) traverse to the observed PPAs.  The fit is
simultaneous to both modes, and the fitted PPA trajectory is shown by
a pair of magenta curves.  The primary mode track is seen at a PPA of
+20\degr\ near the leading edge of the plot and rotates negatively
through about 160\degr\ to +41\degr\ on the far right of the plot.
The longitude origin of the plot is then taken at the inflection
point, and the PPA here is about --58\degr.  This fitted track seems
very plausible in terms of the array of 1.5-GHz PPA ``patches'', but
it is important to resolve whether it is both unique and correct.  Of
course, we have had to omit the non-RVM PPAs around --5\degr\
longitude from the fitting, and we clearly see pairs of other
``patches'' on the conal edges of the profile.  We explored, for
instance, whether the core PPAs from about --9 to 0\degr\ might
represent the same mode as the main conal ``patch'' at 5-6\degr, and
we found not surprisingly that the total PPA traverse is not large
enough to fit these under any conditions.  So, our conclusion is that
most of the core power represents one mode, {whereas the --5\degr\
region} and the brighter emission under the conal outriders the other.

The RVM fit to the 1.5-GHz PPA cannot constrain $\alpha$ and $\beta$
values due to the large correlation of about 98\% between them (see
for e.g.  Everett \& Weisberg 2001).  However, the fitted slope
$R_{PA}$ [=$\sin\alpha/\sin\beta$] of --39$\pm$3 \degr/\degr\ and PPA
at the inflection point of --58\degr$\pm$5\degr\ are well determined,
and the longitude origin is taken at this point in the plots.  The
other way of estimating $\alpha$ comes from the core-width measurement
as was done in ET VI. Using this $\alpha$ (see section~\ref{sec6} and
~\ref{sec11}) as initial value to the fits we obtained $\alpha$ and
$\beta$ values of 125\degr $\pm$19\degr\ and --1.2\degr$\pm$0.5\degr,
where the errors correspond to the fit errors.  We were unable to
carry out a similar fitting for the 4.6-GHz observation---as is pretty
obvious from the nature of the polarized profile.  However, we were
able to ``anchor'' the lower frequency solution on its prominent
leading and trailing PPA ``patches'' and the PPA curve is plotted as
in the lower frequency profile.

That the fitted PPA inflection point falls so late under the core
feature initially surprised us, but we then recalled that we have seen
this behavior under core features in faster pulsars.  With a rotation
period of 359 ms, we suspect that this is due to
aberration/retardation (hereafter A/R) which we will explore further
below.

\begin{figure}
\begin{center}
\begin{tabular}{c}
\mbox{\includegraphics[height=105mm,width=75mm,angle=-90.]{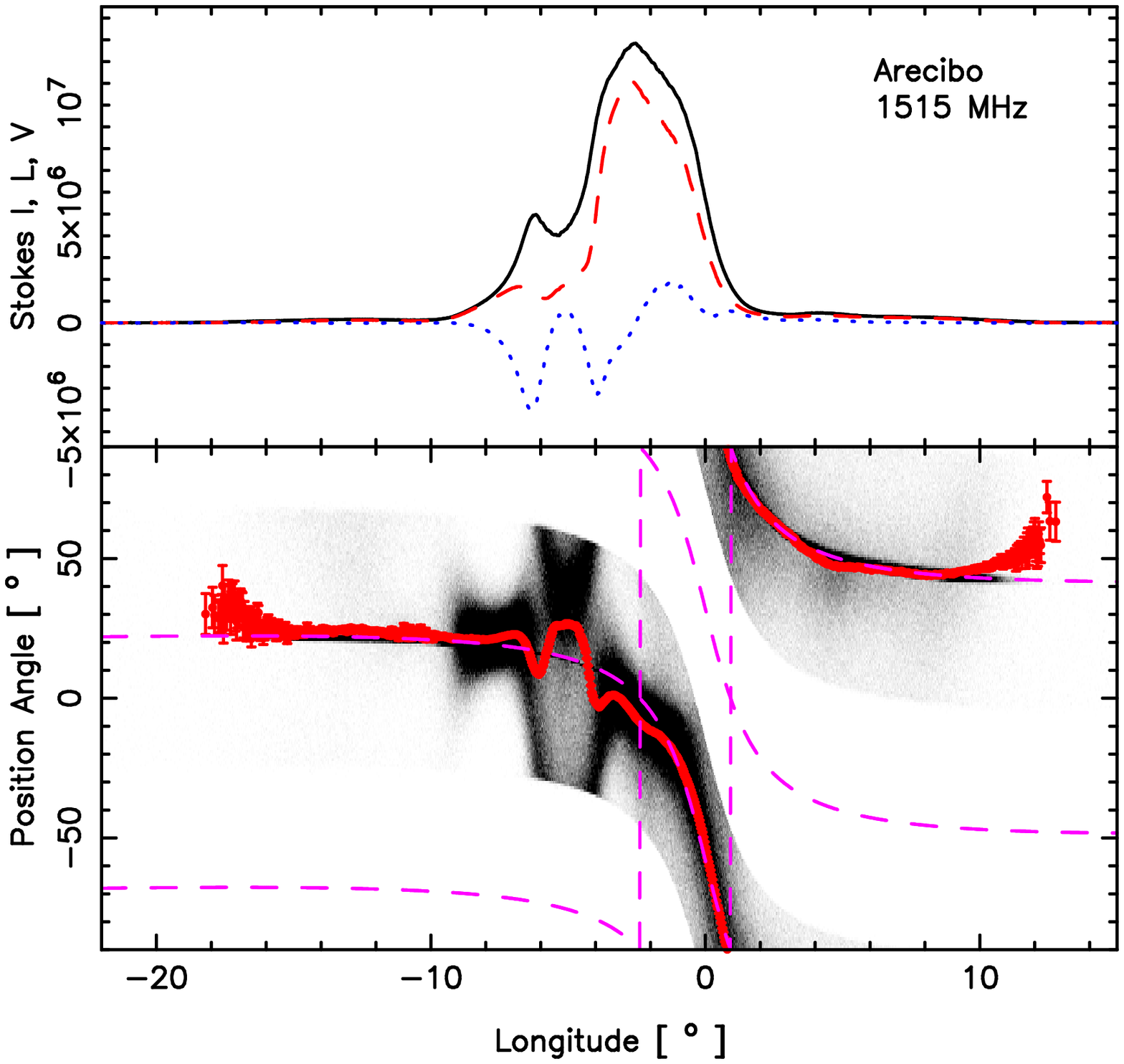}}\\
\mbox{\includegraphics[height=105mm,width=75mm,angle=-90.]{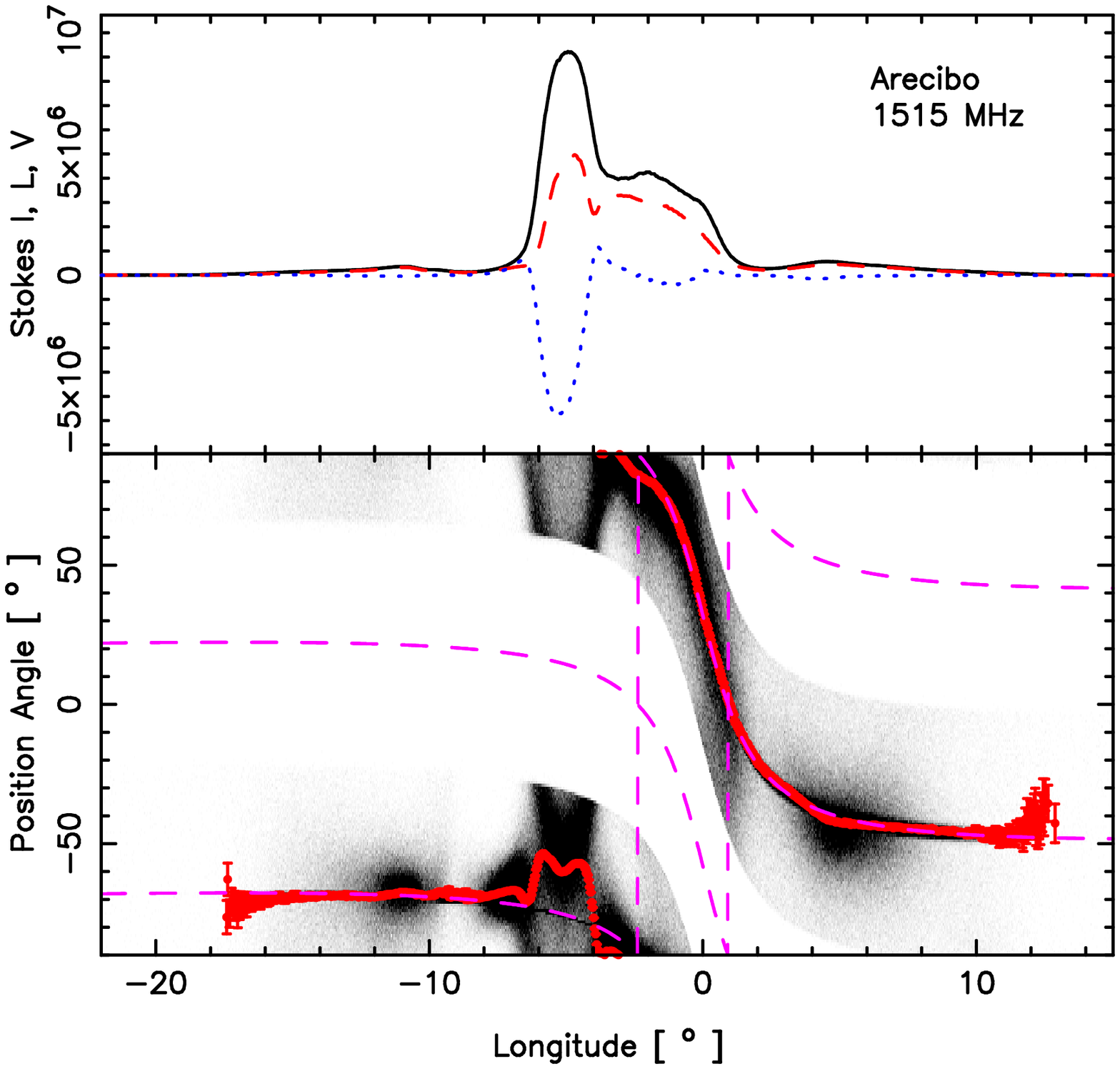}}\\
\end{tabular}
\caption{Polarization profiles and PPA distributions after Figure~\ref{fig1} for two-way 
polarization-mode PS segregations of the 1.5-GHz observation per
Deshpande \& Rankin (2001); primary mode (top) and secondary mode
(bottom).}
\label{fig3}
\end{center}
\end{figure}

Similarly, we were perplexed by the orientation of the
inflection-point PPA ($PA_0$) as one of us had recently interpreted it
as --87\degr\ (Rankin 2015) on the basis of the Johnston \etal's
(2005) average polarimetry.  Surely this gives an object lesson
regarding the difficulties of interpreting average PPA traverses
without appeal to any pulse-sequence polarimetry.  Here we find a
$PA_0$= --58\degr$\pm$5\degr, and we can use this value to estimate
the orientation of the pulsar's rotation axis on the sky so as to then
determine how the radiation is polarized with respect to the emitting
magnetic field direction.  Pulsar proper motions have been found to be
largely parallel to a pulsar's rotation axis (Rankin 2015), and that
of B1933+16 was determined by VLBI to be $PA_v=$+176(0)\degr.  A
similar value was given by Johnston \etal\ as an improved estimate
over that determined earlier by Hobbs \etal\ (2004).  Therefore, the
alignment angle $\Psi = PA_v- PA_0$ is computed --4 -- (--58) =
+54\degr$\pm$5\degr, modulo 180\degr.  Although this value is far from
orthogonal, we will interpret it as corresponding to the extraordinary
(X) propagation mode.

Indeed, the compact distribution of $\Psi$ values (\eg Rankin 2015) is
surprising as there are a number of effects that would tend to
misalign a pulsar's rotation and proper motion direction even if the
natal ``kicks'' are initially aligned (\eg Johnston
\etal\ 2005). Noutsos \etal\ (2012) in particular note the effects of binary disruption 
which might impose a large velocity component orthogonal to the
rotation direction.  We know little about B1933+16's specific origins
or natal binary system, but it is worth noting that there are four
known pulsars within a degree or so of it and at about the same
dispersion measure.  So, the canted $\Psi$ value above should not
immediately be taken as evidence against our association of its
primary core emission with the X mode.  We clearly have much yet to
learn both about why $\Psi$ values tend to close alignment and why
they depart from said.

\section{Intensity Fractionation of the Total Polarized Profiles}
\label{sec4}
Fig.~\ref{fig2} gives intensity-fractionated average profiles for
the two observations.  The 1.5-GHz display in the upper plot exhibits
rather little change, showing that the pulsar's emission is very
steady from pulse to pulse.  There is not enough intensity variation
in its pulses to show large differences.  This said, we can see a
subtle but consistent pattern at both frequencies: the leading portion
of the core feature becomes slightly but progressively weaker relative
to the trailing portion with increased intensity.  Note also that
while the aggregate linear power in the trailing portion of the
1.5-GHz profile depends little on intensity, the leading portion shows
an increase.

\section{Segregating the Polarization Modes}
\label{sec5}
In order to better study the physical emission properties, we next
segregate the modal emission in the 1.5-GHz PS into two sequences.
The procedures for the operations were discussed in Deshpande \&
Rankin (2001; Appendix), and here we first consider the two-way
segregation, which implicitly assumes that the depolarization is modal
in origin, and thus the procedure partially repolarizes the modal
sequences.  Average profiles corresponding to the two modes are given
in Fig.~\ref{fig3}, where the primary polarization mode (PPM) is
plotted in the upper panels and the secondary (SPM) in the lower ones.
Dashed (magenta) RVM curves corresponding to the OPMs are indicated,
and the respective regions of polarized samples contributing to the
average profiles (exceeding a 2-$\sigma$ threshold) clearly shown.

\begin{figure}
\begin{center}
\begin{tabular}{c}
\mbox{\includegraphics[height=105mm,width=75mm,angle=-90.]{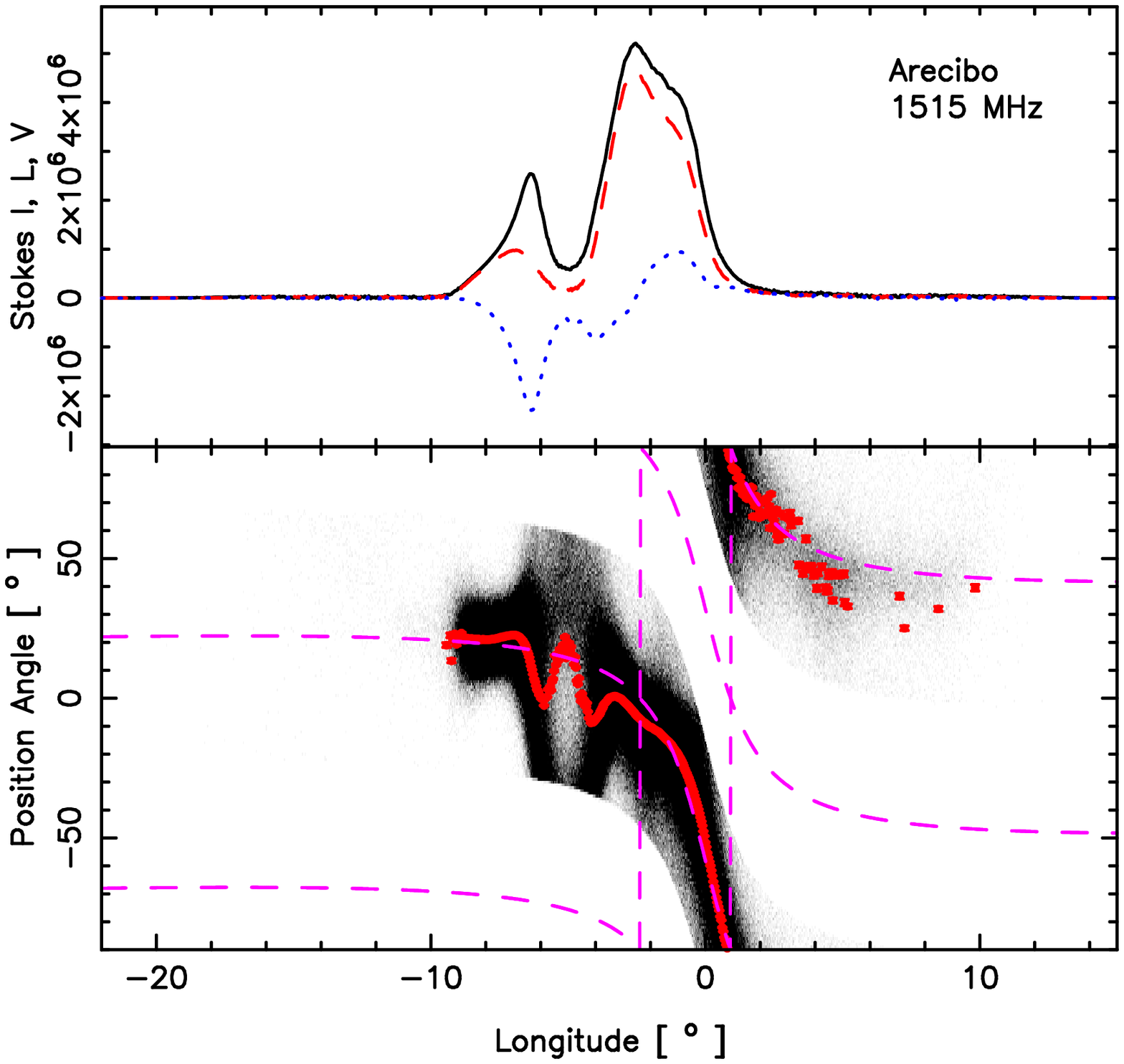}}\\
\mbox{\includegraphics[height=105mm,width=75mm,angle=-90.]{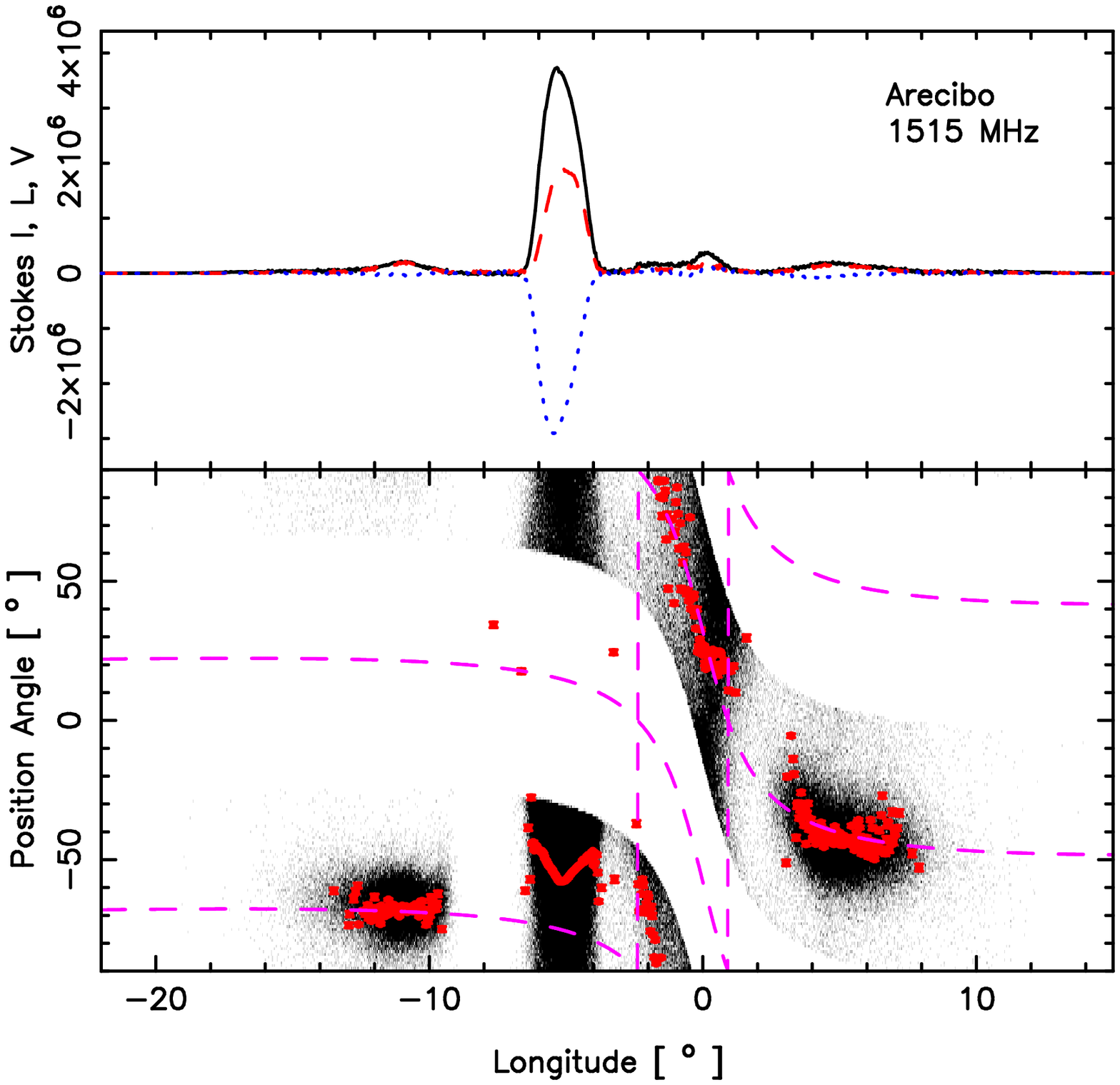}}\\
\mbox{\includegraphics[height=105mm,width=75mm,angle=-90.]{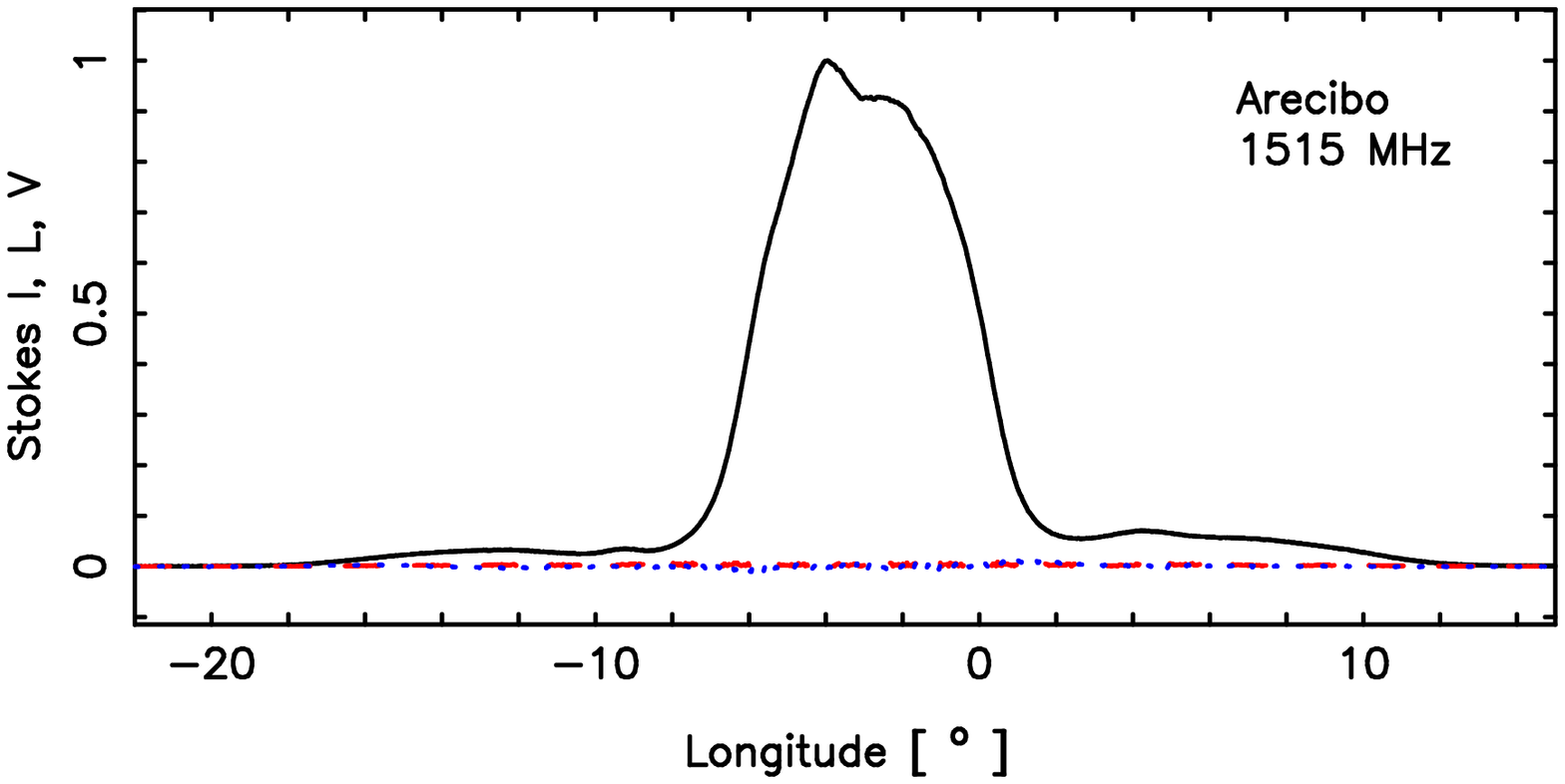}}\\
\end{tabular}
\caption{Polarization profiles and PPA distributions after Figure~\ref{fig1} for three-way 
polarization-mode PS segregations of the 1.5-GHz observation per
Deshpande \& Rankin (2001); primary mode (top), secondary mode
(middle) and unpolarized (bottom).  PPA samples larger than 5$\sigma$
in the linear polarization are plotted in the histogram.}
\label{fig4}
\end{center}
\end{figure}

This polarization-modal segregation shows that the trailing portion of
the core feature plays the strongest role in the PPM and the leading
in the SPM. We also see that most of the conal power is in the SPM.
The two-way modal partial profiles in Fig.~\ref{fig3} are repolarized
following the assumption that the total profile is depolarized due
only to the effects of the two orthogonal polarization modes ;
however, the three-way modal segregations in Fig.~\ref{fig4} are
polarized similarly on their own, suggesting that the assumption is
not wrong.

This all said, it is the large negative `U'-shaped traverse of the PPA
in the 1.5-GHz profile at --5\degr\ that draws the eye.  This feature
is much more prominent in this observation than any other previous one
because of its high resolution---and thus minimal mode mixing within
individual samples.  Many other average profiles show only a ``bump''
at this point (\eg see Johnston \etal\ 2005).  We here see that the
source of this abrupt shift in the average PPA is an aberrant
``patch'' of PPAs at about --50\degr, which falls close to, but
apparently just above, the SPM modal track.  Moreover, the adjacent
regions are unusually prominent here just because of the
non-orthogonality of the --50\degr\ power, as linear polarization
mixing within even these highly resolved samples tends to result in
systematically progressive PPAs.

This interesting configuration is prominent in all of our analyses in
different ways: in the 1.5-GHz total average profile of
Fig.~\ref{fig1}, one can see that the depolarization around --5\degr\
longitude results from PPAs of almost every value.  At 4.6-GHz the
situation is less clear because there are so many fewer qualifying PPA
values.  In the two-way segregations of Fig.~\ref{fig3}, the
repolarization code produces an artifact primarily in the SPM.  Only
in the upper two panels of the three-way segregations do we see a
fairly clear indication of the actual PPA populations here.

The three-way segregations thus give the clearest picture of the
polarization-modal structure of the profile in the --5\degr\ region:
the SPM power in the middle display fits precisely into the region
where PPM power is missing in the top display.  Weak positive circular
polarization is associated with the trailing part of the core;
however, the leading SPM power in the middle display is almost
completely negatively circularly polarized.

\begin{figure*}
\begin{center}
\includegraphics[height=\textwidth,angle=-90.]{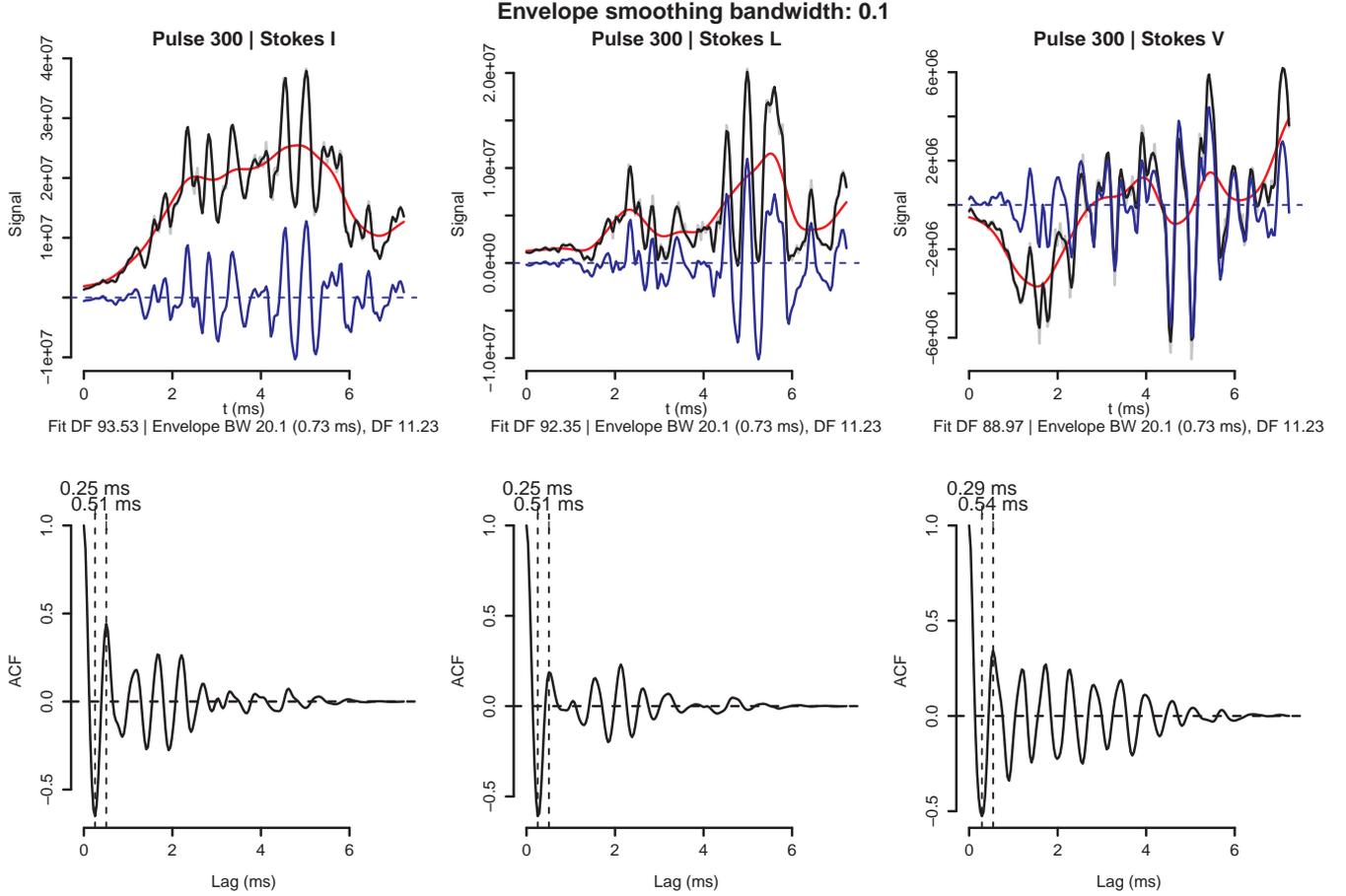}
\caption{\label{fig5-1} Three examples of our analysis methods for extracting subpulse 
timescales in the 1.5-GHz core emission, and the three columns
correspond to Stokes $I,$ $L$ and $V$.  This is similar to MAR15's
fig. 5, but here only the smoothing bandwidth $h=0.1$ is shown.  The
grey lines in the uppermost panels show the subpulse amplitudes, the
fits to them (black lines), the envelopes (red lines), and the
microstructure signal (blue lines), computed as the difference between
the subpulse amplitude and envelope.  The second panel shows the ACF
of the microstructure signal. For a detailed description of the
technique refer to MAR15.}
\end{center}
\end{figure*}

\begin{figure*}
\begin{center}
\includegraphics[height=\textwidth,angle=-90.]{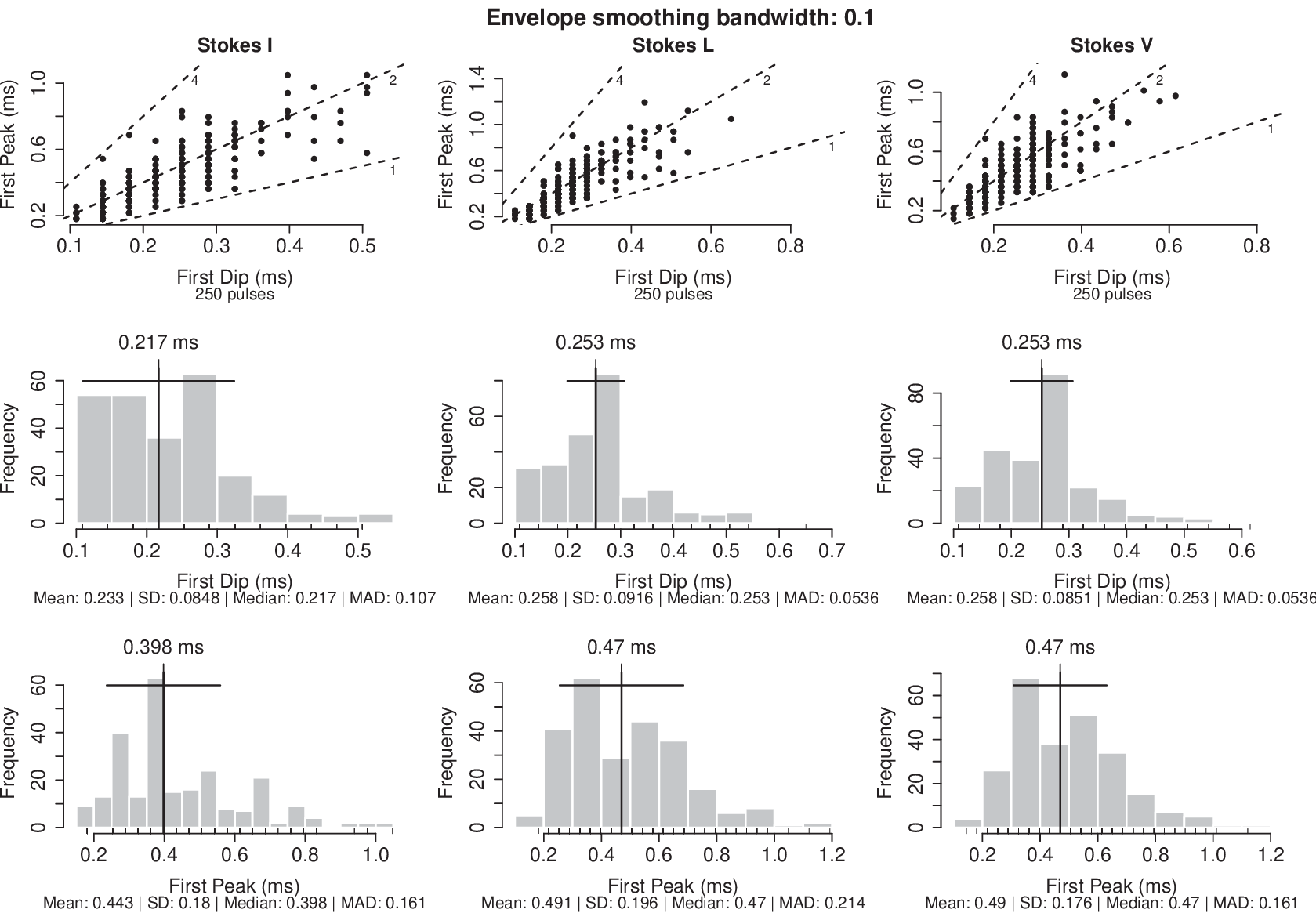}
\caption{\label{fig5-2} The lower three panels are histograms of the estimated time scales 
of the microstructure signals and similar to fig. 6 of MAR15. The
upper plots of these three panels show the relation between the first
dip and the first peak in the ACF, the middle shows the distribution
of the first dip in the ACF and the bottom panel shows the
distribution of the first peak in the ACF which is $P_{\mu}$.  For a
detailed description of the technique refer to MAR15.}
\end{center}
\end{figure*}

\section{Interpreting the Emission Geometry}
\label{sec6}
Although the 1.5-GHz polarization display in Fig.~\ref{fig1} shows a
PPA distribution of unusual complexity (and one is initially at pains
to understand what the 4.6-GHz PPA ``patches'' have to do with the RVM
fit at all!), we believe the RVM fitting to the PPAs in the 1.5-GHz
observation is correct and consistent in the following ways: First,
exempting the region around longitude --5\degr, we can follow a PPA
trajectory under the core feature that begins near --9\degr, sweeps
strongly negative under the trailing edge of the core and includes the
weaker trailing conal PPA ``patch'' at some +45\degr.  We already
noted that an RVM curve cannot be fitted if the brighter trailing
patch is included.  The RVM fit then gives an inflection near
--58\degr.  Second, note that the bright conal PPA ``patches'' both
fall near the SPM curve.  And third, we see that the PPM inflection
point falls near the midpoint or average of the two SPM conal
``patches'' as it must if the RVM model is to be consistent for both
the core and conal regions.

At 4.6 GHz, we have much less to go on.  If this were the only profile
available, no sound interpretation could be made.  However, we note
that the three conal PPA ``patches'', one leading and the two
trailing, fall at nearly identical positions in the PPA distributions
of Fig.~\ref{fig1} that have been derotated to infinite frequency.  So
it is the core region that is mainly different, but even here the PPA
``patch'' at about --1\degr\ falls at about the same place at the two
frequencies.

The effort in ET VI to interpret B1933+16's geometry confronted the
non-Gaussian shape of the pulsar's central feature and remarkably took
twice the peak-to-trailing edge distance, some 4.3\degr, as the ``core
width'' corresponding to the observed angular diameter of the polar
cap.  $\alpha$, $\beta$ estimates of 72\degr\ and 1.1\degr\ followed
which are trivially different that the 125\degr\ (using the Everett \&
Weisberg convention) determined by the RVM fitting above.  The paper's
discussion passed over in silence how the unresolved leading feature
might be interpreted.

The dramatic increasing prominence of B1933+16's conal outrider pair
leaves little doubt, then, that its high frequency profiles represent
a core/cone configuration, but it is less clear just how to classify
it.  As we have seen in the 1.5-GHz profile, the conal power is very
broadly distributed and does not consist in neat components.  Paper ET
VI then gave B1933+16 as a rare example of a core triple {\bf S$_t$}
pulsar with outer conal outriders, but its closer, better defined
conal components at 4.6 GHz defeat this interpretation. Notice the
very different shapes and extents of the conal components at the two
bands in Fig.~\ref{fig1}.  Although no clear double-cone structure can
be discerned, the conal emission is so broad at 1.5-GHz (extending
over some 30\degr\ with half-power points around 22\deg), that outer
conal power must dominate at this frequency.  Further, broad conal
power can be seen in the recent 775-MHz profile of Han \etal\ (2009),
such that the pulsar could be regarded as having a triple {\bf T} or
even five-component {\bf M} profile.  However, these sometimes
``fuzzy'' 1-GHz profile classes all reflect and illuminate the
underlying core/double-cone structure of pulsar radio emission beams.

In any case, using the best determined values of $\alpha$, $\beta$ and
$R_{PA}$ above, the observed angular diameter of the polar cap would
be 5.0\degr, and the roughly 16\degr\ outside half-power width of the
conal outriders at 4.6 GHz definitely identifies them as having the
dimensions of an inner conal beam.  Thus using this profile-width
information together with $R_{PA}$ and the quantitative geometry of
Paper ET VI permits us to confirm that the fitted alpha and beta
values were close to correct.

\section{Microstructure properties} 
\label{sec7}
The bright single pulses of PSR B1933+16 in conjunction with the high
time resolution made possible by the Arecibo observations provide an
ideal context for studying the properties of microstructures in this
pulsar.  Close visual inspection of single pulses at both 1.5 and 4.5
GHz reveals that the bright core component is rich in small timescale
fluctuations in all the Stokes parameters akin to the broad
microstructures recently studied by Mitra, Arjunwadkar \& Rankin
(2015, MAR15 hereafter).  The top panels of Fig.~\ref{fig5-1}
and \ref{fig5-2} show the analysis steps of a single 1.5-GHz pulse as
a typical example of microstructure.  A quantitative analysis of the
pulsar's microstructure properties as carried out by MAR15 was
performed on the strong core component, and the periodicities obtained
for the smoothing parameter $h=0.1$ were $P_{\mu}^{I} \sim 0.4
\pm 0.2, P_{\mu}^{L} \sim 0.47 \pm 0.2, P_{\mu}^{V} \sim 0.47 \pm 0.2$ msec, respectively.  
MAR15 found a relation for $P_{\mu}$ that increases with increasing
pulsar period $P$ for $h=0.1$ as $P_\mu (msec) = 1.3 \times 10^{-3} P
+ 0.04$, predicting a value of about 500 $\mu$sec for PSR B1933+16,
which is in very good agreement with the estimated $P_{\mu}$. The
single pulses at 4.5 GHz were relatively weaker and the microstructure
analysis could only be done reliably for Stokes $I$ which yielded
$P_{\mu}^{I} = 0.35\pm 0.16$ msec for $h=0.1$.

\begin{figure}
\begin{center}
\begin{tabular}{c}
\mbox{\includegraphics[height=75mm,width=75mm,angle=-0.]{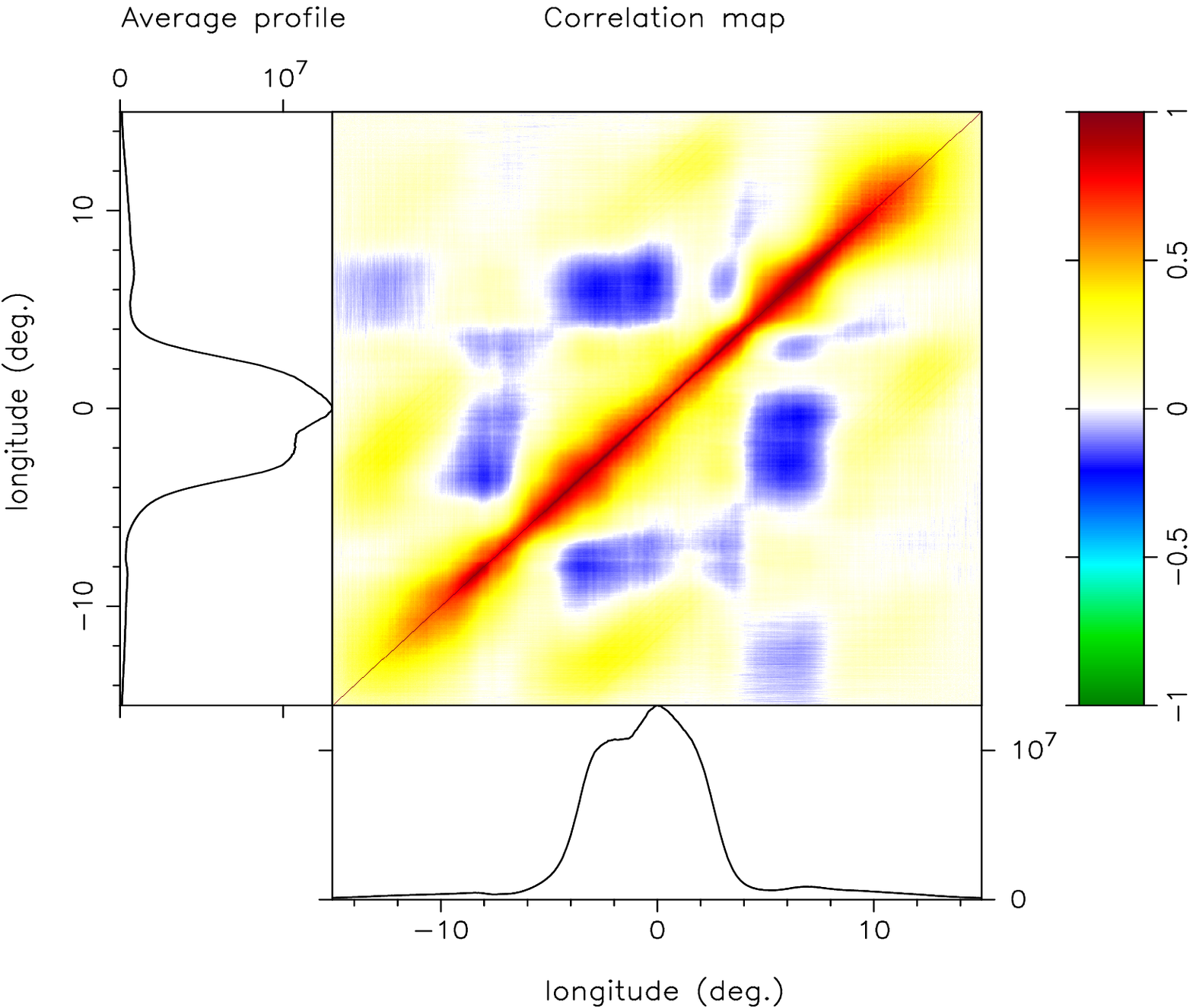}}\\
\mbox{\includegraphics[height=75mm,width=75mm,angle=-0.]{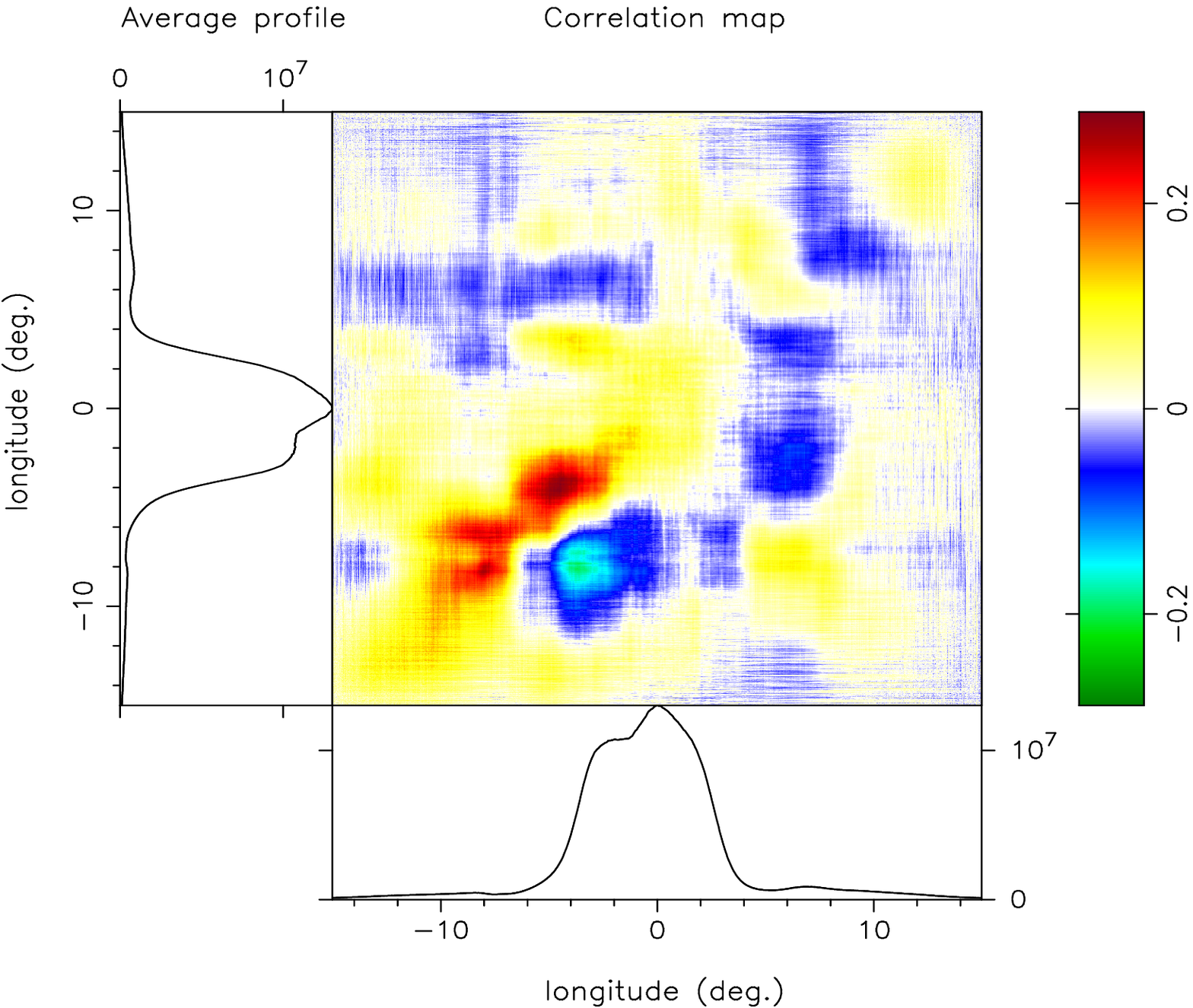}}\\
\end{tabular}
\caption{Longitude-longitude correlation displays for zero delay (top) and a one-pulse delay 
(bottom).  The regions on each side of the diagonal are identical in
the first plot, and in the second the upper region gives delay +1 and
the lower delay --1.  Side and lower panels show the average profile
for reference.  The 3-sigma error in the correlations is about 4\%.}
\label{fig6}
\end{center}
\end{figure}

\begin{figure}
\begin{center}
\mbox{\includegraphics[height=75mm,width=75mm,angle=-0.]{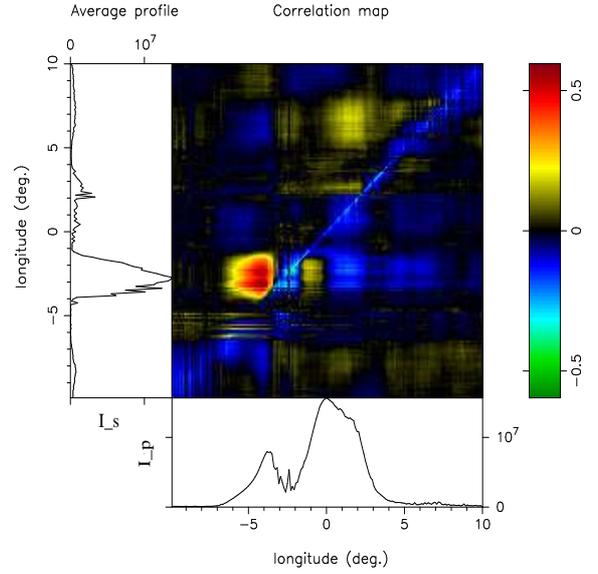}}
\caption{Longitude-longitude cross-correlation display at zero delay between the three-way 
separated primary and secondary OPM power.  Side and lower panels show
the average model profiles (as in Figure~\ref{fig4}) for reference.  The
3-sigma error in the correlations is again about 4\%.}
\label{fig7}
\end{center}
\end{figure}

One of the reasons that motivated the microstructure study was that
PSR B1933+16 is the only truly bright core dominated pulsar in the
Arecibo sky, and no published study of microstructure seems to
specifically address bright core emission. The comprehensive Arecibo
study of broad microstructure in MAR15 treats almost all
cone-dominated stars.  However, we find a striking similarity between
the timescale observed for B1933+16's core component and those for
bright conal emission.

\begin{table}
\caption[]{A/R radio emission heights for pulsar PSR B1933+16, \\
computed as $r = -c P/4$ $\phi_c(\degr)/360$}
    \begin{tabular}{cccccc}
\hline
{\bf Component} & $\nu_{obs}$ & $\phi_l$ & $\phi_t$ & $\phi_c$ & $r$ \\ 
                & (GHz)&  (\degr) & (\degr) & (\degr) & (km) \\
\hline
Core       & 4.5  & --5.4 & 0.16 & --2.78 & 207 \\
Inner Cone & 4.5  &--10.0 & 3.9 & --3.05 & 227 \\  
Outer edge & 4.5  &--14.8 & 11.4 &  --3.4  & 253 \\
           &      &      &      &       &     \\
Core       & 1.5  & --6.1 & 0.00 & --3.05 & 227 \\
Inner Cone & 1.5  &--10.9  & 4.5  & --3.2  & 238 \\  
Outer edge & 1.5  &--16.8 & 11.6 &  --2.6  & 193 \\
           &      &      &      &       &     \\
\hline
    \end{tabular}
\label{tab1}
\end{table}

\section{Conal and Core Emission Characteristics} 
\label{sec8}
Few opportunities exist to study the properties of the conal emission
in core-single {\bf S$_t$} pulsars because the conal outriders appear
only at high frequency where the overall emission already tends to be
weak relative to meter wavelengths.  So it is interesting to carry out
some single-pulse polarimetric analyses on the conal portions of
B1933+16's 1.5-GHz profile.

In order to study the dynamics of the component structure we computed
a longitude-longitude correlation diagram at zero delay, and this is
shown in the top panel of Fig.~\ref{fig6}.  Along the diagonal of
complete correlation, several regions are seen that correspond to the
leading conal, core and trailing conal parts of the profile.  The core
feature shows as a rectangle of weak positive correlation (yellow)
bounded by weak negative (blue).  The conal regions show similar
regions of correlation and interestingly the trailing conal region
seems to bifurcate into an inner and outer region as discussed above.

The lower display in Fig.~\ref{fig6} computes the correlation at
one-pulse delay, and note that the region above the diagonal is no
longer a mirror of that below because the upper portion gives the
correlations at a delay of +1 pulse and the lower at --1.  The region
of large positive correlation at about --4\degr\ longitude falls on
the diagonal and the correlation here is perceptible to delays as long
as 5-10 pulses.  The area of large negative correlation associates the
leading part of the core with the leading conal component, and this
changes greatly with increasing delay.

We also carried out fluctuation spectra on the leading and trailing
conal regions, but these show no strong periodicities.  We see no hint
of intensity-dependent effects as in the core.  The
longitude-longitude correlations do hint at subpulse motion, but not
in a very regular manner as in some slower pulsars.  It is not very
surprising that we see no regular conal modulation, as such ``drift''
modulation seems almost never to occur in stars rotating as rapidly as
B1933+16.

Returning to core properties, Fig.~\ref{fig7} displays the
correlation between the 3-way separated primary and secondary modal
power.  The two different modal profiles that earlier appeared in
Fig.~\ref{fig4} are shown in the side and bottom panels for reference.
First note the bright positive correlation in the early part of the
core feature around --4\degr\ that represents the modal fractions of
power in this region fluctuating together.  Also important are the
weaker regions of positive and negative correlation showing how the
later primary-mode core radiation is coupled dynamically to the
earlier power in the secondary mode.

\section{Aberration/Retardation and Geometrical Emission Heights} 
\label{sec9}
Pulsar B1933+16 exhibits a core-cone geometry.  The prominent conal
peaks are seen clearly at 4.5 GHz adjacent to the central core
component.  At 1.5 GHz the low level conal emission extends far on
both the leading and trailing sides of the profile, however no clear
outer conal peaks are discernible.  Our analysis of the segregated
modes exhibits the largely RVM character of the PPA traverse, and this
prompts us to estimate the emission heights ($r$) for both the core
and conal radiation using the A/R method.

In this method the height $r$ can be determined according to the A/R
shift $\Delta t$=$4r/c$ [=$\Delta\phi P/360\degr$], where $P$ is the
pulsar period in seconds and $\Delta\phi$ is the longitude difference
in degrees between the center of the core or cone emission obtained
from the total intensity profiles and the steepest gradient point of
the PPA traverse (\ie Blaskiewicz \etal\ 1991; Mitra \& Li 2004; Dyks
2008).  $\Delta\phi$ requires determination of the profile center in
terms of the longitude difference between a trailing feature $\phi_t$
and its leading counterpart $\phi_t$, such that $\Delta\phi = \phi_c
- \phi_0$ where $\phi_c = \phi_l + (\phi_t-\phi_l)/2$ and $\phi_0$ is
the ``fiducial'' longitude at the steepest gradient point.  Several
assumptions and systematic errors are associated with $\phi_l$,
$\phi_t$ and $\phi_0$ measurements as discussed in Mitra \& Li
(2004). Also, the A/R method usually assumes an RVM PPA traverse from
a single emission height.  Aggregate emission from several heights
could compromise interpretation of the SG point.  Our purpose here,
however, is to estimate emission heights which are small compared to
the light-cylinder radius.

We apply the A/R method first to B1933+16's conal emission.  At 1.5
GHz the peak location of the inner conal emission based on the average
profile was not easily discernible.  Rather in some low S/N
intensity-segregated profiles the peaks of the cone were visible, and
we used these locations to identify the center of the cone.  The
situation is relatively simple at 4.5 GHz where the conal peaks are
prominent.  We use the profiles in Fig.~\ref{fig1} which are plotted
in a manner such that the longitude origin is taken at $\phi_0$ and
estimate $\phi_l$ and $\phi_t$ for the conal peaks of the inner cone.
Additionally we also measure $\phi_l$ and $\phi_t$ at the extreme
outer edges of the profile at an intensity level of five times that of
the noise level.  These profile measures are given in
Table~\ref{tab1}, including the estimated emission heights.  The
errors in the $\phi_l$ and $\phi_t$ measurements are much smaller than
the systematic errors in this analysis.  Typically we find the conal
emission heights to be about 200 km above the neutron star surface.

Next we proceed to determine the core emission height using A/R.  We
estimate $\phi_l$ and $\phi_t$ using the half power points on either
side of the core feature for both the frequencies.  It is interesting
to see that the midpoint $\phi_c$ falls close to the zero-crossing
point of the sign changing circular. The resulting emission heights
that we obtain are again about 200 km (see Table~\ref{tab1})---very
similar to the conal emission heights.  This is a remarkable result.
For the first time we have successfully been able to apply the A/R
tools to estimate the physical emission heights for both the core and
conal radio emission regions within the profile.

\section{Frequency-Independent OPM Emission} 
\label{sec10}
In general PPA measurements reflect the orientation of the linear polarization electric 
vector with respect to a reference direction.  In our case we measure the PPAs at 1.5 
and 4.5 GHz in an absolute sense (Fig.~\ref{fig1}), which essentially means that the 
PPA values at these two frequencies represent the position angle of the linear 
polarization with respect to celestial North as the emission detaches from the pulsar 
magnetosphere.  As a result, comparison of the PPAs across the profile allows us to 
directly investigate the frequency evolution of the direction of the linear polarization 
for a broad frequency range.

For PSR B1933+16 the PPA comparison is not straightforward due to the
complex nature of the polarization distributions (see \S\ref{sec3}).
In the central core region between --8\degr\ and 0\degr, the PPA
distributions appear to be wildly different at the two frequencies.
Particularly the PPAs near --5\degr\ longitude are concentrated around
--50\degr\ at 1.5 GHz and about --30\degr\ at 4.5 GHz.  The rest of
the PPA distribution earlier than about --8\degr\ and later than
0\degr\ longitude correspond to the conal emission and shows
remarkable similarity between the two frequencies.  The PPA ``patch''
at longitude --10\degr\ is about PPA --60\degr\ and that at longitude
+5\degr\ about --45\degr---and these are virtually invariant with
frequency.  Also the RVM fit that was obtained using at 1.5-GHz PPA
traverse (as shown in Fig.~\ref{fig1}) fits the PPAs at 4.5 GHz in the
conal regions well.

We found an even lower frequency polarization measurement for PSR
B1933+16 at 774 MHz by Han \etal\ (2009).  Although the paper does not
explicitly state that their PPA measurements were calibrated in an
absolute sense, the lead author confirms that they were so (Han 2016).
Their PPAs below the leading conal component fall at about --65\degr\
and those of the trailing at about --50\degr, so they compare quite
closely with our values at 1.4 and 4.5 GHz.  Of course, this situation
was constructed by measuring the Faraday rotation and derotating the
PPAs to infinite frequency in order to achieve the ``absolute''
calibration.  However, it is important here to reverse the question in
order to ask what the broad band PPA alignment indicates about the
propagation of radiation through the pulsar magnetosphere.

\begin{table*}
\caption[]{Summary of A/R core and conal emission heights as in Table~\ref{tab1} for several pulsars in the literature.  
}
    \begin{tabular}{cccccccccc}
\hline
Pulsar   & P & $\dot{P}$ & {\bf Component} & $\nu_{obs}$ & $\phi_l$ & $\phi_t$ & $\phi_c$ & $r$ & Ref \\ 
         &(s)&(10$^{-15}$ s/s)&                 & (GHz)&  (\degr) & (\degr)  & (\degr)  & (km)                                   &    \\
\hline
B0329+54 & 0.714 & 2.05  & Core       & 0.3  &               &              & --1.5 $\pm$ 0.2        & 223$\pm$30 &MRG07 \\
         &       &           & Inner peaks& 0.3  &--6.2$\pm$0.2   &5.0$\pm$0.2&--0.7$\pm$0.2          & 104$\pm$30 &     \\
         &       &           & Outer peaks& 0.3  &--13.8$\pm$0.2    &10.0$\pm$0.2&--1.9$\pm$0.2          & 282$\pm$30 &     \\
         &       &           &            &      &               &              &                       &            &     \\
B0355+54 & 0.156 & 4.4   & Core       & 0.3  &--15.0 $\pm$ 0.5&--8.0 $\pm$0.5 & --11.5$\pm$0.5         & 373$\pm$15 &MR11  \\
         &       &           & Outer peaks&      &--39.7 $\pm$ 0.5& --9.3 $\pm$ 0.4& --15.2$\pm$0.6         & 494$\pm$19 &    \\
         &       &           &            &      &               &              &                       &            &    \\
B0450+55 & 0.340 & 2.37  & Core       & 0.3  &--14.5 $\pm$ 0.5&--7.5 $\pm$0.5 & --11.0$\pm$0.5         & 779$\pm$35 &MR11  \\
         &       &           & Outer peaks& 0.3  &−18.5 $\pm$ 0.2&10.8 $\pm$0.2 & --3.9 $\pm$0.2         & 274$\pm$12 &     \\
         &       &           &            &      &               &              &                       &            &     \\
B1237+25 & 1.382 & 9.06  & Core       & 0.3  &               &              & --0.4$\pm$0.1          & 117$\pm$30 &ERM14 \\
         &       &           & Inner peaks&      & --4.0$\pm$0.2  &3.3$\pm$0.2   &--0.35$\pm$0.2          &101$\pm$50  &     \\
         &       &           & Outer peaks&      & --6.4$\pm$0.2  &5.1$\pm$0.2   &--0.6$\pm$0.2           &172$\pm$50  &     \\

\hline
    \end{tabular}
\label{tab2}
References: MRG07, Mitra, Rankin \& Gupta (2007); MR11, Mitra \& Rankin (2011);  ERM13, Smith, Rankin \& Mitra (2013).
\end{table*}

\begin{table*}
\caption[]{Summary of plasma properties for core and conal emission. The frequency $\nu_{B}$ is 
the cyclotron frequency estimated for $\gamma_s = 200$, frequency
$\nu_p$ estimated for $\gamma_s = 200$ and two values of $\kappa =
100, 10^4$, the cyclotron frequency estimated for two values of
$\Gamma_s = 300, 600$.  The radius of the light cylinder is given by
$R_{lc} = c P/2\pi$ and the height of the radio emission $r$ in terms
of light cylinder distance is given by $R = r/R_{lc}$. The quantities
$\Psi = PA_v- PA_{\circ}$, measured for the primary $\Psi_{PPM}$ and
secondary $\Psi_{SPM}$ polarization modes, are obtained from Rankin
(2015).}  \begin{tabular}{cccccccccccc}
\hline
Pulsar   & P  & $\dot{P}$ & {\bf Component} & $\nu_{obs}$ & $\nu_B$ &  $\nu_p$  & $\nu_{cr}$ & $R$ & R$_{lc}$ & $\Psi_{PPM}$ &  $\Psi_{SPM}$ \\
         &(s) &(10$^{-15}$ s/s)&                 & (GHz)& (GHz)   & (GHz)     & (GHz)      &          & (km)     &($^{\circ}$)                 &($^{\circ}$)   \\
\hline
B1933+16 & 0.358 & 6.0   & Core       & 4.5  & 4676 &20$-$200 &0.5$-$4.3 & 0.01 & 17093 &+112(5)&+22(5) \\
         &       &           & Inner Cone & 4.5  & 3545 &17$-$174 &0.4$-$4.2 & 0.01 &       &       &       \\
         &       &           & Outer edge & 4.5  & 2561 &15$-$148 &0.5$-$3.9 & 0.01 &       &       &       \\
         &       &           &            &      &      &         &          &      &       &       &       \\
         &       &           & Core       & 1.5  & 3545 &17$-$174 &0.4$-$4.2 & 0.01 &       &       &       \\
         &       &           & Inner Cone & 1.5  & 3076 &16$-$162 &0.5$-$4.1 & 0.01 &       &       &       \\
         &       &           & Outer edge & 1.5  & 5769 &22$-$222 &0.5$-$4.5 & 0.01 &       &       &       \\
         &       &           &            &      &      &         &          &      &       &       &        \\
B0355+54 & 0.156 & 4.4   & Core       & 0.3  &  451 &9.4$-$94.4&0.6$-$4.9&0.05&7448     &+89(5) &--1(5)      \\
         &       &           & Outer peaks&      &  195 &6.1$-$62  &0.5$-$4.2&0.07&         &       &     \\
         &       &           &            &      &      &          &         &    &         &       &    \\
B0450+55 & 0.340 & 2.37  & Core       & 0.3  &   54 & 2.2$-$22&0.3$-$2.3 & 0.05 & 16233&+86(16)&--4(16)  \\
         &       &           & Outer peaks& 0.3  & 1234 & 10$-$105&0.5$-$3.9 & 0.02 &      &      &         \\
         &       &           &            &      &               &           &     &       &      &     \\
B0329+54 & 0.714 & 2.05  & Core       & 0.3  &3087 & 11$-$115& 0.4$-$3& 0.006 & 34090 &+99(4)&+9(4)  \\
         &       &           & Inner peaks& 0.3  &30438& 36$-$362 &0.5$-$4.3& 0.003&       &      &          \\
         &       &           & Outer peaks& 0.3  &1526 & 8$-$81   &0.3$-$2.6& 0.008&       &      &          \\
         &       &           &            &      &               &           &     &       &      &        \\
B1237+25 & 1.382 & 9.06  & Core       & 0.3  &19910& 21$-$210& 0.4$-$3.0 &0.002&65890  &+143(4)&+53(4)  \\
         &       &           & Inner peaks& 0.3  &30612& 26$-$261& 0.4$-$3.1 &0.002&       &       &           \\
         &       &           & Outer peaks& 0.3  &6198 & 11$-$117& 0.3$-$2.5 &0.002&       &       &          \\
\hline
    \end{tabular}
\label{tab3}
\end{table*}

\section{On the Structure of Core Components}
\label{sec11}
The obvious composite structure of B1933+16's core component has long
been perplexing.  An idea had developed that core components would be
Gaussian shaped, however those we have studied closely are not, and
perhaps the poor resolution of many older observations had contributed
to this incorrect view.  In any case we must now attempt to interpret
the origins of this structure

The two parts of B1933+16's core component are clearly demarcated by
their differing circular and linear polarization in Figs.~\ref{fig1}
and \ref{fig2}.  The trailing half of the feature has positive
circular and substantial linear, whereas the leading half is linearly
depolarized and shows negative circular.  In earlier work we at some
points tried to understand this two part structure in terms of a
missing ``notch'', but we now see strong evidence that the leading and
trailing parts of the core have very different polarizations and
dynamics.

Paper ET VI interpreted the trailing part of the core as corresponding
to the full width of the polar cap.  The core width there was measured
as twice the peak to trailing half power point or some 4.3\deg,
suggesting a nearly orthogonal value of $\alpha$ (77\degr).  The full
core feature width suggested a less orthogonal geometry which appeared
incompatible quantitatively with the steep $R_{PA}$.  The issue
remains here: we fitted Gaussians to the pulsar's 1.5-GHz profile,
finding that the central core region was well fitted by two Gaussians,
the trailing one with a halfwidth of about 4.5-5\degr, and a leading
one some 3\degr\ earlier of about 2\degr\ halfwidth and half the
amplitude.  This analysis is then compatible with the earlier
assumption that the trailing region is associated with the full polar
cap emission and that the leading feature is ``something extra''.

This line of interpretation is also compatible with the more complete,
probably X-mode linear polarization of the trailing region---and seen
in many other core features (ET XI)---and the complex, depolarized
largely O-mode emission in the earlier part.  The polarization mixing
around --5\degr\ longitude results in the strongly diverted average
PPA traverse and coincides with the peak of the leading fitted
``extra'' Gaussian above.  Both polarization modes are clearly present
in this region and mix both to produce the linear depolarization and
the splay of different polarization states seen in individual samples.
Further, we have seen that higher resolution exhibits the diversion of
the average PPA in this region ever more strongly, so the ``extra''
emission in this narrow region is probably O mode such that its
``patch'' of PPAs in Fig.~\ref{fig1} (upper) would fall on the modal
RVM track if our resolution was even higher.

The distinction between the trailing core region and the ``extra''
leading emission is also seen in the contrasting forms of the
mode-segregated profiles.  We can also see that something peculiar is
happening dynamically at --5\degr\ longitude in the persistent
correlation at 1-pulse delay (Fig.~\ref{fig6} lower), and this effect
is further clarified in Fig.~\ref{fig7} where strong positive
correlation is indicated between the power of the two modes in this
region.  However, it is important to note that the peak falls well off
the diagonal of the diagram, indicating correlation between the modal
power at different longitudes.

Our earlier pulse-sequence polarimetric studies of core emission in
pulsars B0329+54 (Mitra \etal\ 2007) and B1237+25 (Smith \etal\ 2013)
also exhibited core features of two parts, marked by distinct leading
and trailing regions of modal emission and hands of circular
polarization.  Our analyses of these two pulsars went in somewhat
different directions because their cores were modulated by
intensity-dependent A/R, whereas B1933+16's emission is much steadier.
A further difference was that the ``extra'' leading emission had
smaller relative intensity, so that the core width could be directly
interpreted geometrically.

\section{Summary}
\label{sec:summary}
Using high resolution polarimetric Arecibo observations of pulsar
B1933+16 at 1.5 and 4.6 GHz, we have conducted a comprehensive
analysis of this brightest core dominated pulsar in the Arecibo sky.
Here we summarize the main results W of our analysis:
\begin{itemize}
	\item With the exception of one region near --5\degr\ longitude, the 1.5-GHz PPA 
	traverse can be well fitted by an RVM curve.  This shows that most of the core 
	power is in the primary mode and most of the conal in the secondary.  
	\item The 4.6-GHz core polarization is much more difficult to interpret, whereas 
	the conal PPAs are nearly identical to their counterparts at 1.5 GHz.  Imposing 
	the 1.5-GHz RVM fit on the 4.6-GHz PPA provides useful insight on how the 
	latter should be interpreted.  
	\item High resolution analysis reveals the apparently non-RVM region of emission 
	at about --5\degr\ longitude with unprecedented clarity.  
	\item The 1.5-GHz RVM fit gives $\alpha$ and $\beta$ values of 125\degr and --1.2\degr, 
	respectively, with large errors due to the nearly complete correlation between them. 
	The resulting PPA at the inflection point is --58\degr$\pm$5\degr, where we have taken 
	the longitude origin in our various plots. See Fig.~\ref{fig1}.  Efforts were made to 
	interpret the several different core and conal modal ``patches'' in different ways, and 
	we believe the above fit properly represents the pulsar's actual emission geometry.  
	\item Such a ``fiducial'' PPA of --58\degr\ together with the measured  $PA_v$ direction of 
	+176(0)\degr\ indicates polarization orientation +54\degr$\pm$5\degr.  Despite this  
	measured cant from 90\degr, we believe that it probably still identifies the extraordinary 
	(X) propagation mode.
	\item An intensity-fractionation analysis showed only weak differences in the several 
	populations of pulses in B1933+16, in large part because the pulsar emission tends 
	to be very steady from pulse to pulse. See Fig.~\ref{fig2}.  
	\item However, B1933+16's core structure is similar to that seen before in B0329+54 
	and B1237+25---that is, it shows a double modal structure with both opposite senses 
	of circular polarization and orthogonal linear polarization.  In these other pulsars, the 
	two parts are strongly correlated dynamically, but the connection is less apparent 
	in B1933+16 due to its steady intensity from pulse to pulse. Fig.~\ref{fig7}, however, 
	shows the highly correlated mixed-mode power early and its bridge to the later X-mode 
	emission.
	\item The 1.5-GHz modal emission was segregated using both the two- and three-way 
	techniques as seen in Figs.~\ref{fig3} and \ref{fig4}.  These show that the later parts of 
	the core are dominated by X-mode emission, whereas the earlier parts represent the 
	O mode.  
	\item Analysis of broad microstructures in the emission of B1933+16's core component 
	shows that their timescales are nearly identical with the similar largely conal emissions 
	studied earlier----in this case some 500 $\mu$sec.  See Fig.~\ref{fig5-1} \& \ref{fig5-2}.
	\item Longitude-longitude correlations and fluctuation-spectral analyses of B1933+16's 
	1.5-GHz emission show distinct conal and core regions.  The former again show a mix 
	of inner and outer conal radiation.  Fluctuation spectra provide no signs of periodicity.    
	\item Aberration/retardation analyses provide physical emission height measurements 
	for B1933+16's conal emission and unusually here for its core radiation as well.  These 
	computations in Table~\ref{tab1} suggest that both the conal and core radiation stems 
	from an average height of around 200 km above the neutron-star surface.  
	\item Conal linear polarization in B1933+16 exhibits very similar absolute PPAs 
	between 0.77 and 4.6 GHz after derotating them according to the small interstellar 
	Faraday rotation exhibited by the pulsar.  
\end{itemize}

\section{Theoretical Interpretation and Discussion}
\label{sec:discussion}
The enormous brightness temperatures of pulsar radio emission demand a
coherent plasma radiation mechanism.  Relating physical theories of
coherent radio emission to the observations requires above all else
one crucial datum: the location of the regions within the
magnetosphere where the core and conal radio emission are emitted. The
A/R method for determining radio emission heights is physically
grounded and little model dependent, but it has been successfully
applied more to the conal emission, where the RVM can more usually be
fitted to conal PPA traverses and the SG points accurately determined.
Cores have too often presented complex traverses for which it could be
doubted whether an RVM fit was appropriate or even possible.  Some
earlier efforts were made to measure A/R heights for cores (Mitra \&
Li 2004), but the results have remained inconclusive and in some cases
seemed to show the wrong sense.

In an effort to resolve important questions about cores---and their
emission heights in particular---we have embarked on a systematic
study of pulsars with bright core components as well as prominent
conal emission.  This paper is the third in a series that began with
pulsar B0329+54 (Mitra, Rankin \& Gupta 2007) and then studied
B1237+25 (Smith, Rankin \& Mitra 2013).  Each of the pulsars exhibits
distinct conal and central core emission components along with a
complex PPA traverse.  By using high quality single pulse polarimetry
and PPA mode segregation techniques, we were able to identify the
underlying RVM as we did earlier for B0329+54.

This in turn permitted us to estimate the core and conal A/R shifts
that are here summarized in Table~\ref{tab2} (refer to notes in the
Table for details about individual measurements).  Amongst these three
pulsars PSR B1933+16 is the fastest and has the best determined A/R
shift.  We searched the literature and found that for two more
pulsars, PSR B0355+54 and PSR B0450+55, the A/R-based core and conal
shifts could be determined with confidence---\ie we could identify
their components clearly as core or conal and then fit the RVM to
their PPA traverses. As seen in Table~\ref{tab2} the conal and core
centers $\Delta \phi_c$ lead the SG point in each case, which in turn
suggests that the core and conal emission arises from a finite heights
above the polar cap.  However, unlike for B1933+16, the core and conal
centers do not coincide, and hence the A/R shift method, using the
simple relation $r=c\Delta\phi P/4 \times 360^{\circ}$, cannot be used
to estimate these emission heights accurately.  There can be a few
reasons for this, for example: the methods employed to measure the
emission centers are not sufficiently accurate (see Mitra \& Li 2004);
the emission patterns are not centered about the dipole axis: or
finally the core and conal emission arises from different heights.
Our aim here, however, is not to find exact emission heights, but to
illustrate that the radiation arises from deep within the
magnetosphere at altitudes of a few hundred kilometers.  In this
spirit we simply estimate the core and conal emission heights $r$ in
Table~\ref{tab2} assuming their centers are shifted by A/R.  However,
the excellent RVM fits to the PPA traverses in all these cases also
implies that the height differences between the core and conal
emission must be relatively small, as large differences can cause
distortions and kinks in the PPA traverse (see Mitra \& Seiradakis
2004).

These measurements, taken together, provide definitive evidence, we
believe, that the core and conal (hence all) radio emission arises in
regions at heights of typically no more than a few hundred kilometers
above the neutron-star surface. .

Although the PPA traverses of these pulsars are complex and heretofore
inscrutable in RVM terms, our techniques of high resolution
polarimetry, intensity fractionation and modal segregation have been
able to reliably identify their underlying RVM traverses.  B1933+16
was the most complex in these terms we have studied, and we began our
work presuming that its traverse must entail a coherent combination of
polarized power to so distort its observed PPA traverse.  However, as
we have seen, this is not the case: the non-RVM distortions in each of
these pulsars are primarily polarization-modal in origin.  The two
orthogonal polarization (PPM and SPM) modes undergo incoherent mixing
of varying intensities and degrees of orthogonality across the pulse
profile---with the mixing occurring on such short timescales that our
high resolution observations can partially resolve them.  These
effects are particularly acute in the core emission, and there also
complicated by intensity-dependent A/R.

Our ability to trace the PPM and the SPM across the profile and to
identify them with the absolute $PA_{\circ}$ together with the $PA_v$
direction further permits us to associate the OPMs with the X and O
propagation modes in the pulsar magnetosphere.  This is then
summarized in Table~\ref{tab3}.  For pulsars B0329+54, B0355+54 and
PSR B0450+55 the association of the PPM with the X and the SPM with
the O mode is very good. This is in general agreement with the
measured $\Psi$ distribution for core dominated pulsars (Rankin 2015)
where the PPM emission for core components is associated with the X
mode.  For PSR B1933+16 the association is less precise and B1237+25's
canted orientation may stem from its position near the North Galactic
pole.\footnote{It remains surprising that $PA_v-PA_0$ for many pulsars
falls close to either 0 or 90\degr.  Excellent polarimetry and careful
analysis has determined $PA_0$ values for these pulsars with
certainty, and the proper-motion directions measured by both VLBI and
timing are in good agreement with each other.  To the extent that
pulsar velocities are due to natal supernova ``kicks'', the one
alignment can be understood, but binary disruption would tend to
produce the other.  So we have much to learn both about why the
alignments tend to be close to 0 or 90\degr\ as well as why some show
significant misalignments.}  In any case, the observations now
strongly suggest that the emission emerging from the pulsar
magnetosphere is comprised of the X and O modes.

The observational evidence locating the pulsar radio-emission region
and associating the radiation with the X and O plasma modes have
strengthened our ability to identify the curvature-radiation mechanism
as responsible for exciting coherent radio emission in pulsars (\eg
Melikidze, Gil \& Pataraya 2000, hereafter MGP00; Gil, Luybarski \&
Meikidze 2004, hereafter GLM04; Mitra, Gil \& Melikidze 2009;
Melikidze, Mitra \& Gil 2014, hereafter MMG14).  In most pulsar
emission models the the rotating neutron star is a unipolar inductor
due to the enormous magnetic fields, and a strong electric field is
generated around the star.  In such strong electric and magnetic
fields charged particles can be pulled out from the neutron star
and/or be created by the process of magnetic pair creation.  Finally
the region around the neutron star becomes a charge-separated
magnetosphere which is force free.  The physics of how particles
populate and flow in the pulsar magnetosphere is a matter of intense
research; however, the consensus is now that, in the presence of
sufficient plasma, a relativistic plasma flow can be generated along
the global open dipolar magnetic field [see Spitkovsky (2011) for a
review].  In order to achieve a force-free condition and to maintain
corotation the magnetosphere needs a minimum charge density
(Goldreich \& Julian 1968) of $n_{GJ} = \Omega \cdot B/2 \pi c$
(rotational frequency $\Omega = 2 \pi / P$, where $P$ is the pulsar
period, $B$ is the magnetic field and $c$ is velocity of light).  A
key aspect of the magnetosphere is that enormous electric fields will
be generated in regions depleted below the $n_{GJ}$ value, where the
remaining charges can be accelerated to high energies and more charges
produced in that region.

In order to explain the origin of pulsar radio emission arising
relatively close to the star, a charge-depleted acceleration region
just above the polar cap is found to be necessary.  Such a prototype
region was envisaged as an inner vacuum gap (IVG) by Ruderman \&
Sutherland (1975, RS75).  The gap has strong electric and magnetic
fields and is initially charge starved.  Apparently, the IVG can
eventually discharge as electron-positron pairs in the form of
``sparks'' at specific locations, and corresponding non-stationary
spark-associated relativistic primary particles with Lorentz factors
of $\gamma_p$ can be generated.  This discharge process continues
until the force-free charge density $n_{GJ}$ is reached. The
electron-positron pairs are separated due to the electric field in the
IVG, and one kind of charge streams outward towards the upper
magnetosphere further radiating in the strong magnetic fields and the
resulting photons thereby producing secondary electron-positron plasma
with Lorentz factor $\gamma_s$.  Thus the charge density in the
outflowing secondary plasma $n_s$ gets multiplied by a factor
$\kappa \sim n_{s}/n_{GJ}$ (Sturrock 1971).  The backflowing plasma
then heats the polar cap surface and can generate thermal x-ray
emission.  For the above process to work, the RS75 model assumed a
magnetic curvature radius of 10$^6$ cm in the vacuum gap, which in
fact implies strongly non-dipolar fields (see \eg Gil, Melikidze
\& Mitra 2004).  As a consequence the range for $\gamma_p \sim 10^{6}$, $\nu_s \sim$ few 
hundreds, and $\kappa \sim 100-10^{4}$.  The RS75 model had some shortcomings, however, 
as it could not explain the slower-than-expected subpulse-drifting effect (Deshpande \& Rankin 
2001), and it also predicted higher temperatures for the hot polar caps than were observed.
A refinement of the model by GLM04 suggested that a partially screened IVG can be
realized that can explain these effects.  In any case, the core and conal radio emission are 
generated through the growth of plasma instabilities in the spark-associated relativistically 
flowing secondary plasma.

Knowledge of the location of the pulsar emission and using the above
model, it is possible to compute the various frequency scales involved
in the pulsar emission problem.  Following eqs. (10), (11) and (12) in
MMG14, the cyclotron frequency $\nu_{B}$, and plasma frequency
$\nu_{p}$ at a fractional light cylinder distance $R$ for a pulsar
with period $P$ and $\dot{P} = \dot{P}_{15} \times 10^{-15}$ can be
expressed as $\nu_{B} = 5.2
\times 10^{-2} (1/\gamma_s) \times (\dot{P}_{15}/P^5)^{0.5} R^{-3}$ GHz and $\nu_{p} = 2 \times 
10^{-5} \kappa^{0.5} \sqrt{\gamma_s} (\dot{P}/P^7)^{0.25} R^{-1.5}$
GHz.  In Table~\ref{tab3} we provide the values of $\nu_{B}$ and
$\nu_{p}$ for a reasonable value of $\gamma_s =200$ and two values of
$\kappa = 100 , 10^4$, respectively.  For all cases we find $\nu_{p}
< \nu_{B}$, and further the the observed radio emission $\nu_{obs}
< \nu_{p}$, or in other words the observed frequency of radio emission
is lower than the plasma frequency of the emitting plasma.

A model of pulsar radio emission for $\nu_{obs} < \nu_{p} < \nu_{B}$
has been developed by MGP00.  It is important to notice that the only
known instability that can operate at low radio emission heights
(typically below 10\% of the light cylinder) is the two-stream
instability.  The non-stationary sparking IVG discharge leads to
generation of secondary plasma clouds with slight spreads in their
particle velocities.  The faster and slower velocities of two
successive clouds overlap at the heights of radio emission to trigger
strong Langmuir turbulence in the secondary plasma, which can become
modulationally unstable.  MGP00 demonstrated that nonlinear growth of
the modulational instability can lead to formation of charged solitons
which are capable of emitting curvature radiation in curved magnetic
fields.  The soliton size must be larger than the linear Langmuir
wave, and to maintain coherence the emission should have wavelengths
larger than the soliton size, or in other words $\nu_{obs} < \nu_{p}$
as observed.  The theory suggests that the maximum frequency of the
soliton coherent curvature radiaiton given by eq. (12) in MMG14 is
$\nu_{cr} = 0.8 \times 10^{9} (\Gamma^3/P) R^{-0.5}$ GHz, where
$\Gamma$ is the Lorentz factor of the emitting soliton.  The value of
$\Gamma$ is slightly different from $\gamma_s$, and assuming a two
reasonable values of $\Gamma = 300,600$ we estimate values of
$\nu_{cr}$ as given in Table~\ref{tab1}.  Clearly the values lie in
the observed frequency range---\ie $\nu_{cr} \sim \nu_{obs}$.

Finally we turn our attention to the prediction of the coherent
curvature radiation theory for the X and O modes and their propagation
in the magnetosphere.  We have emphasized in \S\ref{sec10} that in PSR
B1933+16 the linear polarization direction remains unchanged for conal
emission between 0.77 and 4.5 GHz.  In turn, if we invoke the RVM
solution, we can conclude that with respect to the magnetic field line
planes there is no rotation or adiabatic walking of the linearly
polarized conal modal emission in the pulsar magnetosphere.  We note
that several current studies suggest the invariance of the PPAs with
frequency.  One class of investigation involves establishing the
rotation measure variation as a function of pulse phase
(Ramachandran \etal\ 2004; Noutsos \etal\ 2009), and these studies
indicate that the rotation measure across a pulsar's profile is
largely due to the interstellar medium.  Another study by
Karastergiou \& Johnston (2006) compared absolute PPAs between 1.4 and
3.1 GHz and found very little change with frequency.

If the PPA modes are interpreted as the X and O modes, then the
absence of adiabatic walking, at least for the X mode, can be
understood in the framework of the propagation effects predicted by
the coherent curvature radiation theory as shown in a recent work by
MMG14.  This theory demonstrated that for radio emission to be excited
and detach from the magnetosphere below 10\% of the light cylinder
(which is also the case for PSR B1933+16 as shown above), the
refractive index of the emitting plasma for the X-mode propagation is
unity, and hence the X mode can emerge from the plasma preserving its
emitted polarization direction. However the theory has no prediction
for the O mode nor for the existence of circular polarization.  These
appear to be two fundamental problems that need to be resolved in
pulsar emission theories as emphasized in the latter paper.

\section{Conclusion}
In this paper we have established that the core and conal emission lie
at heights of no more than several hundred kilometers, core and conal
emission display similar short timescale microstructures, and their
emission can also be associated with the X and O modes.  These
similarities strongly suggest, as expected, that the core and conal
emission processes have the same physical origin.  However, we have
yet to understand the causes of their individual geometric,
polarization and modulation properties.  We argue that, overall, the
pulsar radio emission mechanism is excited by coherent curvature
radiation.

\section*{Acknowledgments} We thank our referee Jarsalow Dyks for providing constructive comments
which helped in improving the paper. Much of the work was made
possible by support from the NASA Space Grants and from US National
Science Foundation grant 09-68296.  One of us (JMR) also thanks the
Anton Pannekoek Astronomical Institute of the University of Amsterdam
for their support.  Arecibo Observatory is operated by SRI
International under a cooperative agreement with the US National
Science Foundation, and in alliance with SRI, the Ana
G. M\'endez-Universidad Metropolitana, and the Universities Space
Research Association.  We thank Prof. Jinlin Han for sharing their
774-MHz polarimetry of PSR B1933+16. This work made use of the NASA
ADS astronomical data system.

{}
\end{document}